**Quantifying Key Design Factors for Thermal Comfort in Underground Space Through Global Sensitivity Analysis and Machine Learning**

Shisheng Chen[a, b, c], Nyuk Hien Wong[c], Chao Cen[c], Ruohan Xu[c], Lei Xu[d], Zhenjiang Shen[a,b],

Zhigang Wu[a], Jiayan Fu[e], Zhongqi Yu[e, f, g] *

a- School of Architecture and Urban-Rural Planning, Fuzhou University, China 350108.
b- Laboratory of Smart Habitat for Humanity, Fuzhou University, China 350108.
c- Department of the Built and Environment, National University of Singapore, 4 Architecture Drive, Singapore, 117566.
d- Future Cities Lab (FCL) Global, Singapore-ETH Centre, Singapore 138602
e- College of Architecture and Urban Planning, Tongji University, China 200092.
f- Key Laboratory of Ecology and Energy Saving Study of Dense Habitat, Ministry of Education, 200092 China.
g- Tongji Architectural Design (Group) Co., Ltd., China 200092.

* Corresponding author.

E-mail address: yzq_caup@tongji.edu.cn (Z Yu)

**Abstract**:

This study identified the key design factors related to thermal comfort in naturally ventilated underground spaces under high temperature condition (outdoor $T_{max}$= 42.9 °C) in Fuzhou, China. Fuzhou has a humid subtropical climate and is one of the three hottest cities in China in 2024. Daytime measurements indicated reduced air temperature (AT), mean radiant temperature (MRT), wind speed (V), and elevated relative humidity (RH) in the underground space. Physiological equivalent temperature (PET) in the underground was consistently lower during peak hours (hour 8–16), with the maximum difference in PET between pedestrian and underground levels being 11–11.9°C. Higher pedestrian-level PET at L1 was attributed to reduced greenery and shading, and decrement factors indicated greater thermal dampening at L2 (0.197) than L1 (0.308). Sensitivity analysis found MRT was the most influential factor (S1/ST = 0.59–0.72) for aboveground spaces, followed by AT (0.13–0.26). In contrast, underground PET was mainly affected by metabolic rate (MET) (0.63–0.65), followed by RH (0.14–0.20) and V (0.08–0.18). Partial dependence revealed that a 1 met increase in MET raised PET by 1.6 °C, whereas a 1 m/s increase in V reduced PET by 1.5–2.2 °C in the underground space. Due to cooler and more stable thermal conditions, underground spaces have higher tolerance for intensive physical activities. By buffering fluctuations in AT and MRT, underground environments can significantly alleviate heat stress and provide passive cooling shelter during daytime heat waves. Overall, this study provided empirical evidence to underground space design in hot-humid climate offering insights for sustainable urban heat mitigation.

**Key words**: Underground; Thermal comfort; Physiological Equivalent Temperature; Extreme high temperature; Sensitivity Analysis; Machine Learning

**Nomenclature**

| Symbol | Meaning | Symbol | Meaning |
| --- | --- | --- | --- |
| UHI | Urban heat island | RF | Random forest |
| PET | Physiological equivalent temperature | XGB | Xgboost |
| UTCI | Universal thermal climate index | RMSE | Root mean squared error |
| Out_SET* | Outdoor standard effective temperature | SMAPE | Symmetric mean absolute percentage error |
| WBGT | Wet bulb globe temperature | $h_c$ | Convective heat transfer coefficient |
| LAI | Leaf area index | $h_{c,free}$ | Free convective heat transfer coefficient |
| Tmax | Maximum temperature | $h_{c,forced}$ | Forced convective heat transfer coefficient |
| GT | Globe temperature | $\varepsilon$ | Surface emissivity of black-globe |
| AT | Dry-bulb air temperature | $\sigma$ | Stefan-Boltzmann constant |
| WBT | Wet-bulb temperature | D | Diameter of black-globe |
| V | Wind speed | $\mu$ | Decrement factor |
| RH | Relative humidity | $\Delta T_{max,indoor}$ | Maximum air temperature difference at indoor environment |
| MRT | Mean radiant temperature | $\Delta T_{max,outdoor}$ | Maximum air temperature difference at outdoor environment |
| M | Metabolic rate | S1 | First-order sensitivity index |
| $I_{cl}$ | Clothing insulation level | ST | And total-order sensitivity index |
| KS | Kolmogorov-smirnov | $Y_i$ | Observation |
| MLR | Multiple linear regression | $\hat{Y_i}$ | Predicted value |
| Std | Standard deviation | CV | Coefficient of variation |

# 1. Introduction

## 1.1 Extreme high temperature events in urban environment

Urban microclimatic conditions directly affect the comfort and productivity of people in outdoor environments, particularly in densely populated urban canopy areas. As global climate change intensifies, rising outdoor temperatures driven by the urban heat island (UHI) effect have become increasingly severe posing a significant challenge to urban liveability and residents' quality of life. A notable manifestation of this challenge is the increasing frequency and severity of extreme heat events. The investigation of role of heat domes in the unprecedented 2021 heatwave in Western North America discovered that the heat dome contributed over 50% of the observed temperature anomalies (Zhang et al., 2023). It showed that the intensities of hot extremes associated with similar heat domes were increasing faster than background global warming, partly due to soil moisture-atmosphere feedback. Extreme high-temperature events in the China–Pakistan Economic Corridor from 1961 to 2015 were investigated using daily maximum temperature data (Li and Bao, 2023). The results show that the frequency, intensity, and duration of extreme heat events have significantly increased over this period. The generalized Pareto distribution predicted the highest recurrence levels reaching up to 50°C for 100-year return periods. In addition, summer extreme heat events in the Beijing–Tianjin–Hebei region from 1979 to 2020 reported an overall upward trend and significant increases in both the extreme maximum temperature and the number of high-temperature days after the mid-1990s (Liang et al., 2023). Spatially, these events were concentrated in the southern parts of the BTH region with EMT increasing by 1.5–2.0°C in many areas. Another study was conducted to identify the changes in extreme high temperatures and population heat exposure across 289 Chinese cities from 2002 to 2020 using MODIS land surface temperature data (Zhang et al., 2024). It concluded that extreme temperature thresholds increased by 0.43°C per decade during the day and 0.67°C per decade at night, with higher thresholds in developed cities. The number of cities exposed to compound heat extremes rose from 230 to 256, and nearly 42% of cities experienced such extremes for 15 consecutive years or longer, particularly in East and South-Central China.

Recent data (Figure 1) shows major cities across China generally experienced persistent high temperatures in Summer in 2024, especially in the city of Chongqing, Hangzhou, Fuzhou, Wuhan, Nanchang, and Shanghai (China Meteorological Administration, 2024). The highest temperature in these cities reached 42°C and the duration of high-temperature throughout the year reached 77 days significantly exceeding the historical level for the same period. Extreme high temperatures not only increase energy consumption, but also pose a serious threat to public health. Extreme heat adversely affects physical health (e.g., dehydration, heat exhaustion), mental health (e.g., increased anxiety, depression), and behavioural aspects (e.g., reduced outdoor activities, increased violence) (Diallo et al., 2024). Low-income and minority communities are more vulnerable to extreme heat due to factors such as air conditioning and limited green space. Children are more at risk from sudden weather and temperature changes than adults (Yang et al., 2025b). Outdoor workers also face significantly higher exposure to extreme heat compared to indoor residents increasing their risk of heat-related illnesses such as heat stroke, cramps, heat exhaustion, heat rash, loss of productivity and ability to work (Kjellstrom et al., 2009; Yang et al., 2025a). To address the increasing number of extreme heat events, evidence-based mitigation strategies are needed to reduce heat-related health risks and energy threats.

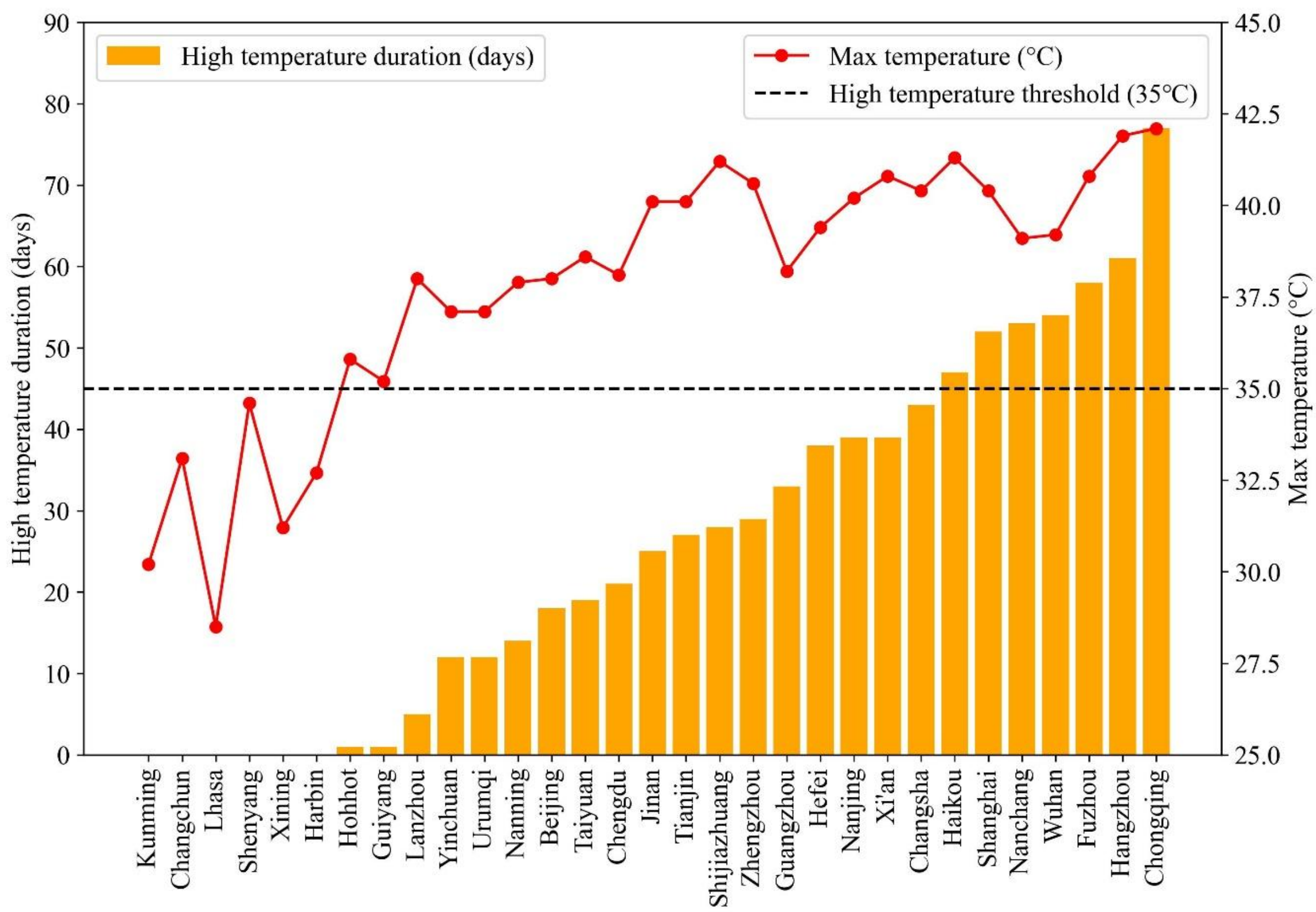

Figure 1: High temperature duration and maximum temperature in major cities across China in 2024. (Data source: China Meteorological Administration (China Meteorological Administration, 2024))

### 1.2 Thermal comfort of underground space

There have been some studies on thermal comfort in underground space. For data collection and evaluation metrics, thermal comfort data typically employed field measurements, questionnaire surveys, and simulation-based modelling. Onsite monitoring and interviews were used to assess PMV and compare the thermal comfort between underground and aboveground spaces (Tan et al., 2018). CFD simulation was integrated with field measured data to evaluate thermal comfort in tropical basements (Wen et al., 2024a). Real-time sensor data was applied to calculate the PMV-PPD in metro environments (Katavoutas et al., 2016). Thermal comfort was inferred from numerical modelling using simulated airflow and temperature in traditional underground buildings (Wen et al., 2023). Another study explored the impact of surrounding rock thermal properties on the thermal comfort of an underground refuge chamber using CFD simulation and field experiments (Jin et al., 2025). Extreme environments such as deep mine evaluated the heat stress and thermal responses combining both field experiments and climate chamber test data (Wang et al., 2025). These approaches guarantee both quantitative rigor and human centric validation of thermal conditions in underground environments. In terms of thermal comfort indices, the indicators for indoor and outdoor environments are clearly defined, and various metrics have been applied to assess underground thermal comfort, including MRT (Yang et al., 2019), SET* (Fang et al., 2025), PET (Chen et al., 2024; Yang et al., 2019; Zhao et al., 2023), Thermal Sensation Vote (Li et al., 2018b; Liu et al., 2025; Yang et al., 2022), and WBGT (Liu et al., 2025). To enhance the thermal comfort in underground spaces, different strategies were proposed including climate-responsive ventilation strategies (Tan et al., 2018), aboveground–underground integration (Wen et al., 2024a), passive pre-cooling devices (Wen et al., 2023), mechanical cooling (Katavoutas et al., 2016), personalized and physiological controls (Wang et al., 2025). These studies advocate a multiscale, climate- responsive, and human centric design approach balancing passive strategies with mechanical systems to ensure thermal comfort in diverse underground environments.

Thermal comfort in underground spaces is climate-sensitive. In hot and humid regions, unique challenges arise different from those in temperate or hot and dry regions, particularly due to the synergetic effects of high humidity and elevated temperatures on thermal comfort. Hot and humid environments are more likely to cause heat-related illnesses (e.g., heat stroke) (Heidari et al., 2020). Underground spaces located at dry climate environments have low humidity level and minimal moisture-related issues, which are mainly concentrated on sensible cooling loads (Li et al., 2018a). A study conducted in Fuzhou, which has a humid subtropical climate, reported that underground spaces failed to maintain thermal comfort due to discomfort caused by high humidity (RH=78.8-80%) despite stable air temperatures (Tan et al., 2018). In contrary, the thermal comfort of underground space can be maintained within the comfort zone (PMV= 0.38-1.21) without active cooling in a temperate climate city like Beijing (RH=44.2-54.7%) (Tan et al., 2018). Similarly, passive ventilation is sufficient in Nanjing (RH 53.5–56.9%), where the underground space can maintain the PMV 1-1.52 without active systems (Tan et al., 2018). However, this does not mean that natural ventilation cannot be achieved in hot and humid climates. In Singapore, a city-state with a hot and humid climate, natural ventilation has been well implemented in the design and construction stage for both commercial and residential buildings (Deng et al., 2023). To achieve effective natural ventilation in underground spaces in hot and humid climates, the key is to integrate aboveground and underground environments through cross ventilation, windcatchers, greenery, sunshades and urban air corridors to balance wind velocity and heat exchange (Wen et al., 2024a, 2023).

### 1.3 Research gap and objectives

In rapidly urbanizing and high-density cities such as Shanghai and Fuzhou in China, available ground space is very limited. Naturally ventilated underground spaces offer a thermally stable microclimate due to aboveground thermal mass, shading from direct solar radiation, and reduced exposure to anthropogenic heat.

Despite the potential, thermal comfort studies of naturally ventilated underground environments are limited. Existing thermal comfort studies on underground spaces have focused either on mechanically ventilated deep environments (e.g., metro stations, refuge chambers) or on passively ventilated spaces under normal summer or moderate conditions, without exposure to extreme heat. In contrast, few studies have continuously monitored the diurnal dynamics of thermal comfort in naturally ventilated shallow underground spaces during heat events in hot-humid climates. Moreover, evidence on how shallow underground microclimates perform relative to adjacent aboveground spaces under the same extreme heat conditions remains largely absent, leaving a gap in scientific design guidance for integrated aboveground and underground development in high-density cities in hot-humid climate. To address this gap, this study applied Sobol sensitivity analysis and machine learning techniques to quantify the relative influence of environmental and design factors on thermal comfort, representing a methodological advancement seldom employed in previous research. Therefore, the objectives of this study were to: (1) Evaluate the diurnal variation in thermal comfort in naturally ventilated shallow underground space for human activity under extreme summer heat in hot-humid climate; (2) Identify the environmental parameters and design factors influencing thermal comfort in underground spaces under extreme heat condition; (3) Provide evidence-based thermal comfort strategies to support integrated aboveground and underground space development in high-density cities.

The remaining structure of this paper is organized as follows. Section 2 describes the overall research design including study areas, environmental monitoring, and calculation of thermal comfort. Section 3 reports the comparative results of environmental conditions and thermal comfort conditions at different locations. Sensitivity analysis is conducted to explore the impact of environmental factors and human metabolism on Physiological Equivalent Temperature (PET) distributions. Section 4 and Section 5 discusses and concludes the environmental conditions and thermal comfort of underground space and its potential as the heat shelter during the extreme heat period.

## 2. Methodology

### 2.1 Research design

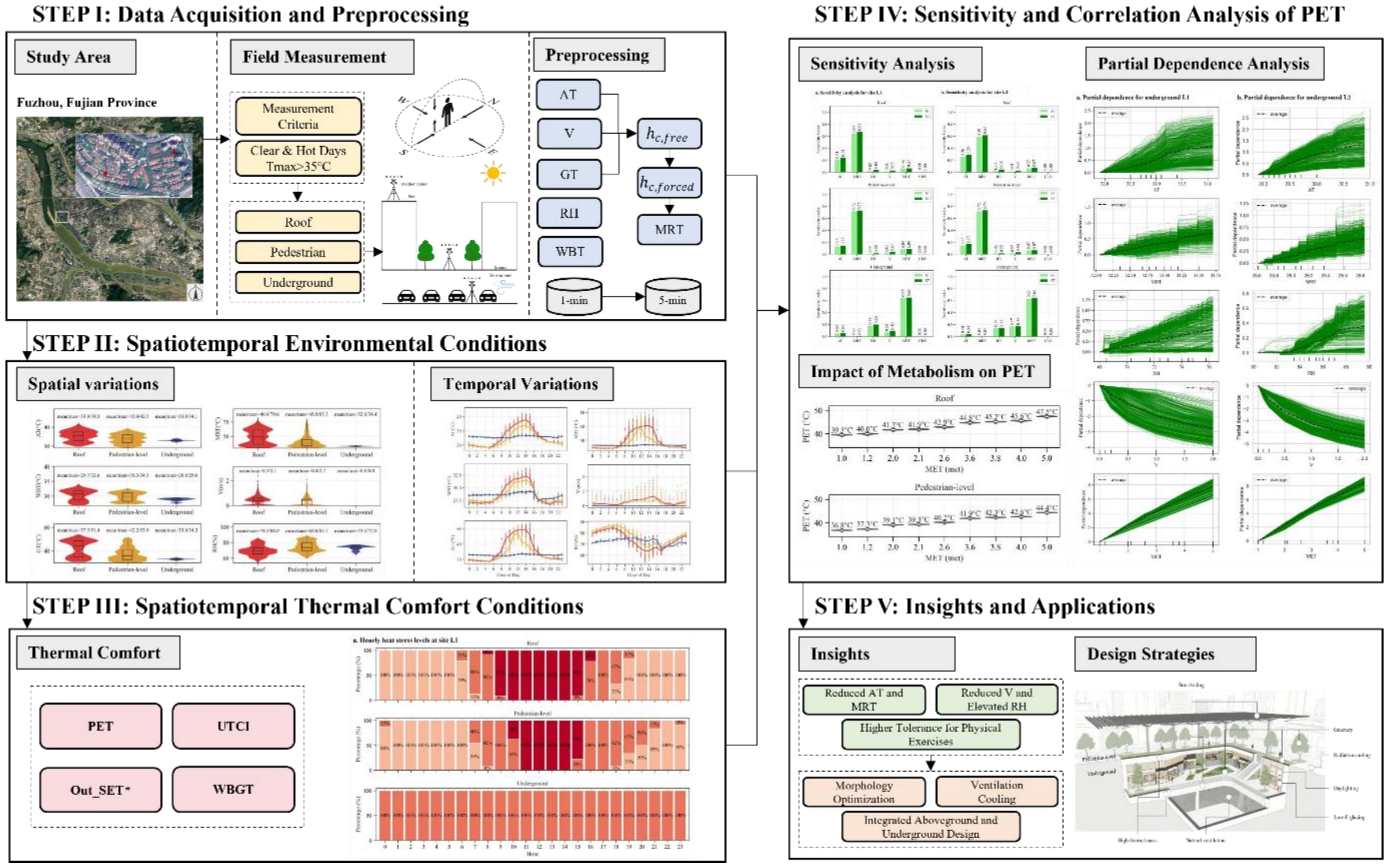

Figure 2: Research design

The research design outlined in the Figure 2 was structured to investigate thermal comfort of different urban layers. The field campaign was designed to address the research gap by targeting extreme heat events (Tmax ≥ 35 °C) in hot-humid climate in a naturally ventilated shallow underground space actively occupied by humans. The test site and measurement period were selected to represent conditions seldom explored in previous studies, which mostly focused on mechanically cooled or deep underground environments under moderate weather condition.

High-resolution microclimate monitoring was conducted including 1-min logging of globe temperature (GT), dry-bulb air temperature (AT), wet-bulb temperature (WBT), wind speed (V), and relative humidity (RH), complemented by simultaneous aboveground reference measurements.

For thermal comfort measurement, this study employs the black-globe thermometer method to measure mean radiant temperature using Kestrel 5400 featured by a 25mm black copper globe. In addition, PET was used to assess the perceived temperature by occupants based on meteorological factors and physiological factors (e.g., metabolic rate, clothing level).

Statistical comparison across urban layers was implemented to identify diurnal patterns and differences in environmental conditions and thermal comfort. The temporal and spatial distributions were computed to understand how thermal comfort varied over time and across different urban locations. To ensure a robust and statistically significant conclusion, bootstrap resampling and uncertainty quantification were conducted on the PET data. Sobol sensitivity analysis and machine learning were employed to quantify the relative influence of environmental and design factors, providing novel insights into the effectiveness of shallow underground space as heat shelter and informing mitigation strategies for urban heat exposure in high-density cities in hot-humid climate.

## 2.2 Study areas and scopes

The study areas (Figure 3) were located on a residential estate in Fuzhou, Fujian Province, China. The environmental monitoring was conducted in two sites (e.g., L1, L2) and three urban layers (e.g., Roof, Pedestrian-level, Underground). Site L1 was surrounded by 6-floor low-rise residential blocks and Site L2 was surrounded by 18-floor mid-rise residential blocks. The data from rooftop measurement point at Site 1 were used as the reference for both two sites because the location was unshaded and exposed to the maximum of solar irradiance. The pedestrian level measurement points were located on the ground floor, and the underground level measurement points were located in the underground parking lots with similar depth and ground connectivity. The facility management has disabled mechanical ventilation in the parking lot, but entrance opening was provided for natural ventilation. After measuring wind speeds at different sites and heights, the maximum wind speed was 3 m/s recorded on the roof and the minimum wind speed was 0 m/s at underground level, all of which were natural convection.

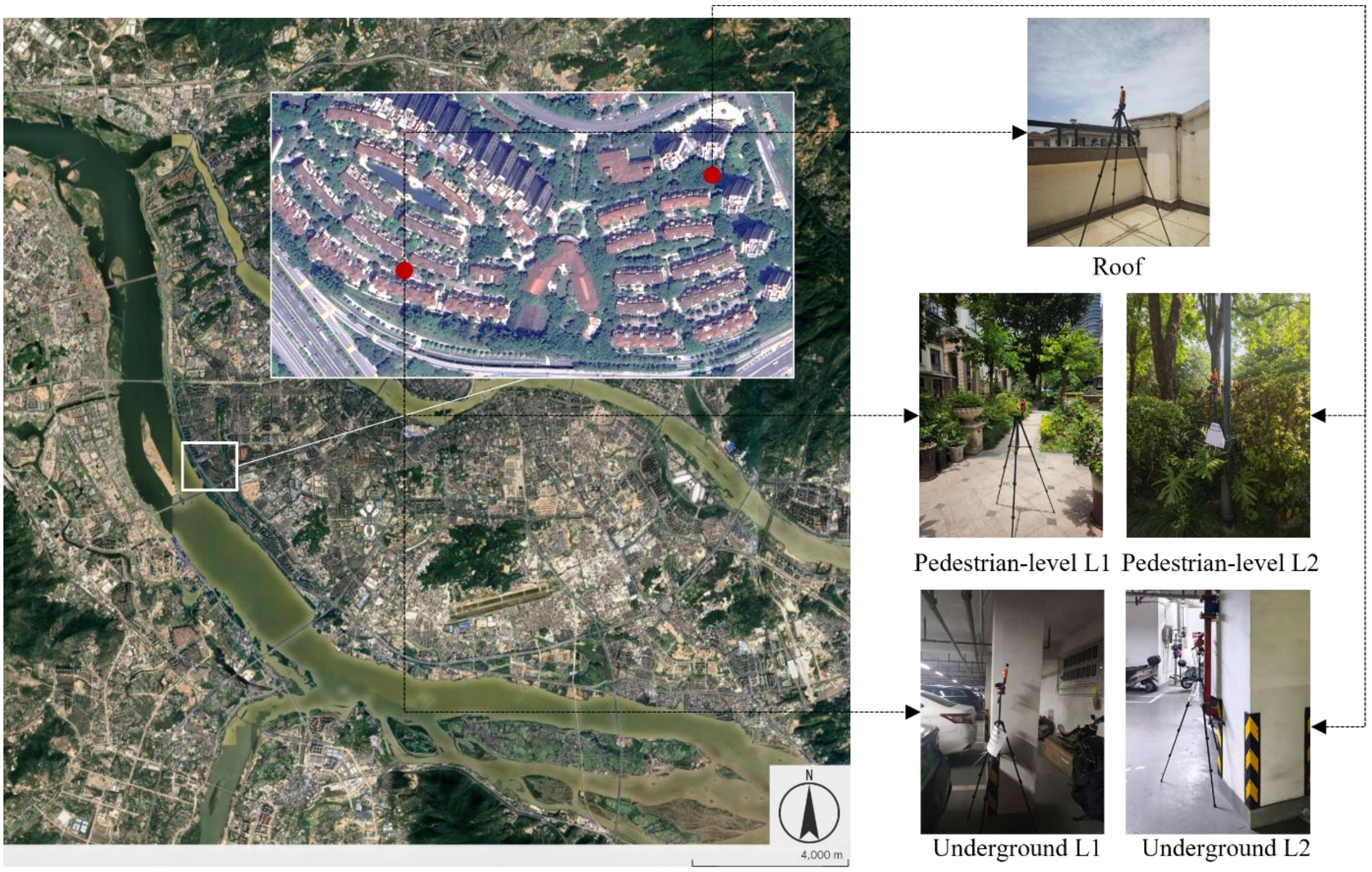

Figure 3: Study areas

According to Köppen climate classification (Beck et al., 2018), Fuzhou has temperate climate with hot summer and no dry season. Fuzhou is one of the three hottest cities in 2024, with 58 high temperature days and a maximum temperature of 40.8°C (China Meteorological Administration, 2024). Based on historical data from 1991 to 2020, Fuzhou experiences mean daily air temperature of 28.9–29.4 °C and mean daily maximum air temperature of 33.8–34.6 °C, accompanied by 6.1–7.2 h of sunshine per day during July and August (China Meteorological Administration, 2020). The peak value of 42.9 °C recorded in this study thus represented an extreme heat event above the typical temperature range in summer. According to the high temperature alerts standards established by the China Meteorological Administration, high temperature warning are categorized into three tiers: yellow warning when the maximum temperature (Tmax) will exceed 35°C in the next 72 hours, orange warning when the Tmax will exceed 37°C in the next 24 hours and red warning when the Tmax will exceed 40°C in the next 24 hours (China Meteorological Administration, 2007). To evaluate the thermal comfort of different spaces under high temperature condition, this study selected clear, sunny days with Tmax > 35°C for experiments. The experiments were conducted from August 7$^{th}$ 2024 to August 13$^{th}$ 2024. GT, AT,

WBT, V and RH were recorded at 1-min frequency using Kestrel 5400. The specifications of environmental sensors are described in Appendix A.

In terms of research scopes, this study focused on the diurnal variation in thermal comfort in naturally ventilated shallow underground spaces within 10 m below ground used for human activity under extreme summer heat (July-August) in hot and humid climate. Due to limited availability of suitable sites, a naturally ventilated underground parking facility was selected as the representative research site. Experiments were conducted with the mechanical ventilation system turned off to simulate the natural ventilation condition.

**2.3 Estimation of mean radiant temperature**

The black-globe thermometer method can be used to estimate the mean radiant temperature (MRT). The black-globe temperature represents the weighted average of the radiation temperature and the ambient temperature. Based on ISO 7726 (International Organization for Standardization, 1998; Teitelbaum et al., 2022), MRT can be calculated based on Equation (1), where the free convection coefficient from Equation (2) or the forced convection coefficient from Equation (3) is applied, whichever is greater.

$$MRT = \sqrt[4]{(GT + 273.15)^4 - \frac{h_c}{\varepsilon\sigma} \times (AT - GT)} - 273.15 \quad (1)$$

$$h_{c,free} = 1.4\sqrt[4]{\frac{AT - GT}{D}} \quad (2)$$

$$h_{c,forced} = 6.3\frac{V_a^{0.6}}{D^{0.4}} \quad (3)$$

Where MRT= mean radiant temperature, °C; AT= dry-bulb air temperature, °C; GT= globe temperature, °C; $h_c$= convective heat transfer coefficient, W/(m²·K); $h_{c,free}$= free convective heat transfer coefficient, W/(m²·K); $h_{c,forced}$= forced convective heat transfer coefficient, W/(m²·K); $\varepsilon$= surface emissivity of black-globe, 0.95; $\sigma$= Stefan-Boltzmann constant, 5.67e-8 W/m²·K⁴; D= diameter of black-globe, 0.15 m.

The convective heat transfer coefficient of globe surface is affected by environmental conditions such as temperature, humidity and wind speed. Therefore, there is certain discrepancy in the MRT estimated using a black-globe thermometer compared to the integral radiation measurements. The integral radiation method calculates the MRT based on the average radiation flux of long-wave and short-wave radiation absorbed by the human body in six directions (Thorsson et al., 2007).

Although MRT measured through integral radiation measurements is considered the ground truth for MRT assessments, substantial improvements can be achieved through 5-minute resampling. Research conducted in Sweden demonstrated there was an improvement in the 5-minute correlation between MRT measured by globe thermometer and MRT obtained through integral radiation measurements under clear weather conditions (Thorsson et al., 2007). In addition, another study revealed the 5-minute correlation between MRT measured by globe thermometer and MRT measured by integral radiation during summer conditions also showed a notable enhancement when compared to 1-minute raw data collected in Harbin (Du et al., 2021). Similar findings were observed in Wuhan, China where 5-minute resampling increased the coefficient of determination ($R^2$) from 0.84 to 0.91 in MRT estimation using globe thermometer (Li et al., 2024). Furthermore, a study in Hongkong also discovered that 5-minute resampling could produce an $R^2$ above 0.9 (Ouyang et al., 2022). These findings underscore the effectiveness of 5-minute resampling to enhance the accuracy of MRT measurements using black-globe thermometer.

## 2.4 Evaluation of thermal comfort

The Universal Thermal Climate Index (UTCI), Outdoor Standard Effective Temperature (Out_SET*), Wet Bulb Globe Temperature (WBGT) and PET are suitable indices widely used for calculating heat stress in outdoor space, which are used to evaluate the thermal comfort across different measurement points in this study. Solar radiation load is applied to the SET calculations for locations (e.g., roof, pedestrian-level) exposed to direct solar radiation based on ISO 7243: 2017 (International Organization for Standardization, 2017). Pythermalcomfort package (Tartarini and Schiavon, 2020) was used to calculate these thermal comfort indices. The input parameters of metabolic rate (M) and clothing insulation level ($I_{cl}$) were first assigned with a fixed M of 1.2 met and $I_{cl}$ of 0.5 clo representing the typical summer attire wearing short-sleeve shirt, thin trousers or skirt, underwear, T-shirt, socks and shoes. For heat stress evaluation, PET values were classified into different heat stress levels according to Table 1.

Table 1: Classification of heat stress based on PET (Matzarakis et al., 1999)

| PET (°C) | Heat Stress Category |
|---|---|
| < 4 | Extreme Cold Stress |
| 4 - 8 | Strong Cold Stress |
| 8 - 13 | Moderate Cold Stress |
| 13 - 18 | Slight Cold Stress |
| 18 - 23 | No Thermal Stress |
| 23 - 29 | Slight Heat Stress |
| 29 - 35 | Moderate Heat Stress |
| 35 - 41 | Strong Heat Stress |
| ≥ 41 | Extreme Heat Stress |

For simulation of the impact of environmental variables and human metabolism on thermal comfort, M and $I_{cl}$ were assigned to the value specified in Table 2 and 0.5 clo representing the typical summer attire from ASHRAE Standard 55 (ASHRAE, 2017).

Table 2: Simulation settings for activities, metabolic rate and clothing insulation

| Activity | M (met) | $I_{cl}$ (clo) |
|---|---|---|
| Seated, quiet | 1 | 0.5 |
| Standing, relaxed | 1.2 | 0.5 |
| Walking 0.9 m/s | 2.0 | 0.5 |
| Walking 1.2 m/s | 2.6 | 0.5 |
| Walking 1.8 m/s | 3.8 | 0.5 |
| Lifting/packing | 2.1 | 0.5 |
| Handling 50 kg bags | 4 | 0.5 |
| Tennis | 3.6 | 0.5 |
| Basketball | 5.0 | 0.5 |

## 2.5 Evaluation of decrement factor

The decrement factor quantifies how much thermal damping occurs as a temperature wave moves through a structure, reflecting the reduction in temperature fluctuation experienced on the opposite side of the material. As shown in Equation (4), a modified decrement factor ($\mu$) was calculated to quantify the thermal mass of underground structure at site L1 and site L2, respectively.

$$\mu = \frac{\Delta T_{max,indoor}}{\Delta T_{max,outdoor}} \quad (4)$$

Where $\Delta T_{max,indoor}$ represents the maximum air temperature difference at indoor environment within 24 hours; $\Delta T_{max,outdoor}$ represents the maximum air temperature difference at outdoor environment within 24 hours. In this study, $\Delta T_{max,indoor}$ and $\Delta T_{max,outdoor}$ were referred to maximum air temperature difference at underground and pedestrian-level, respectively.

**2.6 Statistical and regression analysis**

The descriptive statistics such as mean, standard deviation (Std), coefficient of variation (CV) of environmental variables and thermal comfort indices were reported in Appendix B. The Kolmogorov-Smirnov (KS) test was used to determine whether two distributions were from same distribution. No prior testing of underlying distribution is required for KS test. The null hypothesis assumes two dataset values (e.g., roof & underground) are from the same continuous distribution and is rejected when P-value is less than 0.05. One-way ANOVA was used to evaluate the effect of MET on PET across different locations.

Bootstrap resampling was applied to estimate the underlying population distributions of PET at roof level, pedestrian level, and underground. For each condition (e.g., metabolic rate), the PET values were resampled with replacement to generate 5000 bootstrap replicates of the sample mean. This non-parametric resampling approach allowed for robust estimation of mean and confidence interval in PET providing meaningful statistical comparisons of thermal comfort conditions in these naturally ventilated environments.

Global sensitivity analysis and partial dependence plot were conducted on analytical model and regression model of PET, respectively. In this study, the analytical model referred to the original PET calculation model in pythermalcomfort (Tartarini and Schiavon, 2020), and the PET regression model (Equation (5)) referred to the training model established using experimental data. Since clothing insulation values for typical summer ensembles of male and female do not differ much, the regression model did not simulate the effect of $I_{cl}$ on PET. The candidates of regression model including Multiple Linear Regression (MLR), Random Forest (RF), and XGBoost (XGB) from Scikit-learn Python package (Pedregosa et al., 2011) were applied for the training of experimental data and inference of PET. Machine learning models can capture nonlinear relationships between environmental variables and PET which are suitable for representing results from analytical models with minimal loss of accuracy. The hyperparameter tuning (Table 3) of RF and XGB was conducted through Bayesian optimization using BayesSearchCV from Scikit-optimize Python package (Head et al., 2018).

$$\mathrm{PET} = f(\mathrm{AT}, \mathrm{MRT}, \mathrm{RH}, \mathrm{V}, \mathrm{M}) \quad (5)$$

This research employed root mean squared error (RMSE) and symmetric mean absolute percentage error (SMAPE) as performance metric of regression model, with model performance assessed through 5-fold cross-validation. In contrast to the mean squared error, RMSE (Equation (6)) gives an error measurement in the same units as the dependent variable and imposes a stricter penalty on large errors compared to the mean absolute error. SMAPE (Equation (7)) penalizes over-predictions and under-predictions equally in percentage terms.

$$\mathrm{RMSE} = \sqrt{\frac{1}{\mathrm{n}}\sum_{\mathrm{i}=1}^{\mathrm{n}}\left(\mathrm{Y_i} - \hat{\mathrm{Y}}_{\mathrm{i}}\right)^2} \quad (6)$$

$$\mathrm{SMAPE} = \frac{100}{\mathrm{n}}\sum_{i=1}^{n}\frac{2 \cdot \left|\mathrm{Y_i} - \hat{\mathrm{Y}}_{\mathrm{i}}\right|}{\left|\mathrm{Y_i}\right| + \left|\hat{\mathrm{Y}}_{\mathrm{i}}\right|} \quad (7)$$

where n is the number of observations; $Y_i$ is a given observation; $\hat{Y}_i$ is the predicted value.

Table 3: Hyperparameter optimization

| Model | Hyperparameter | Range |
|---|---|---|
| RF | n_estimators | (50, 200) |
| | max_depth | (3, 20) |
| | min_samples_split | (2, 10) |
| XGB | n_estimators | (50, 200) |
| | max_depth | (3, 15) |
| | learning_rate | (0.01, 0.3, 'log-uniform') |

Partial dependence was used to quantify the correlation between input variables and outcome variable (e.g., MRT on PET) computed based on "sklearn.inspection.PartialDependenceDisplay" from Scikit-learn Python package (Pedregosa et al., 2011).

Global sensitivity analysis was implemented using SALib Python package to quantify the contribution of input variables on the outcome variable (Herman and Usher, 2017). The bounds of input variables in sensitivity analysis are listed in

Table 4. First-order (S1) and total-order (ST) sensitivity index were reported. The S1 quantifies how much a single input variable contributes to the variance in the model output. The ST accounts for both the individual effects and all possible interaction effects involving each input. Sobol method was used for global sensitivity analysis leveraging Monte Carlo sampling to assess the influence of input variables on model outputs.

Table 4: The bounds of input variables in sensitivity analysis

| Site | Measurement point | Variable | Min | Max |
|---|---|---|---|---|
| L1 | Roof | AT (°C) | 28.70 | 42.50 |
| | | MRT (°C) | 27.37 | 83.33 |
| | Pedestrian-level | AT (°C) | 28.90 | 39.30 |
| | | MRT (°C) | 28.68 | 79.61 |
| | Underground | AT (°C) | 30.90 | 34.10 |
| | | MRT (°C) | 30.58 | 34.36 |
| L2 | Roof | AT (°C) | 28.40 | 42.90 |
| | | MRT (°C) | 27.50 | 80.66 |
| | Pedestrian-level | AT (°C) | 28.50 | 40.70 |
| | | MRT (°C) | 28.00 | 84.93 |
| | Underground | AT (°C) | 28.50 | 30.90 |
| | | MRT (°C) | 23.93 | 30.81 |
| L1/L2 | Roof/ Pedestrian-level/ Underground | V (m/s) | 0 | 2 |
| | | RH (%) | 50 | 90 |
| | | M (met) | 1 | 5 |
| | | $I_{cl}$ (clo) | 0.36 | 0.57 |

**3. Results**

**3.1 Spatiotemporal variations of environmental conditions**

To distinguish the diurnal and nocturnal thermal behaviours, Figure 4 and Figure 5 were plotted to illustrate the mean and maximum values of environmental variables at site L1 and site L2 during the day and night, respectively. The underground location generally experienced more stable and cooler conditions across both sites, with less variability in AT, WBT, GT, MRT, and higher RH. As shown in Table B7 (Appendix B), KS test confirmed that there were statistically significant differences in environmental variables between sites L1 and L2 (p-value <0.05), except V in the underground space. At the pedestrian-level, site L2 had slightly lower mean AT, WBT, GT, MRT, and RH, and slightly higher V. In the underground space, site L2 also had lower AT, WBT, GT, and MRT, but higher RH, compared to site L1.

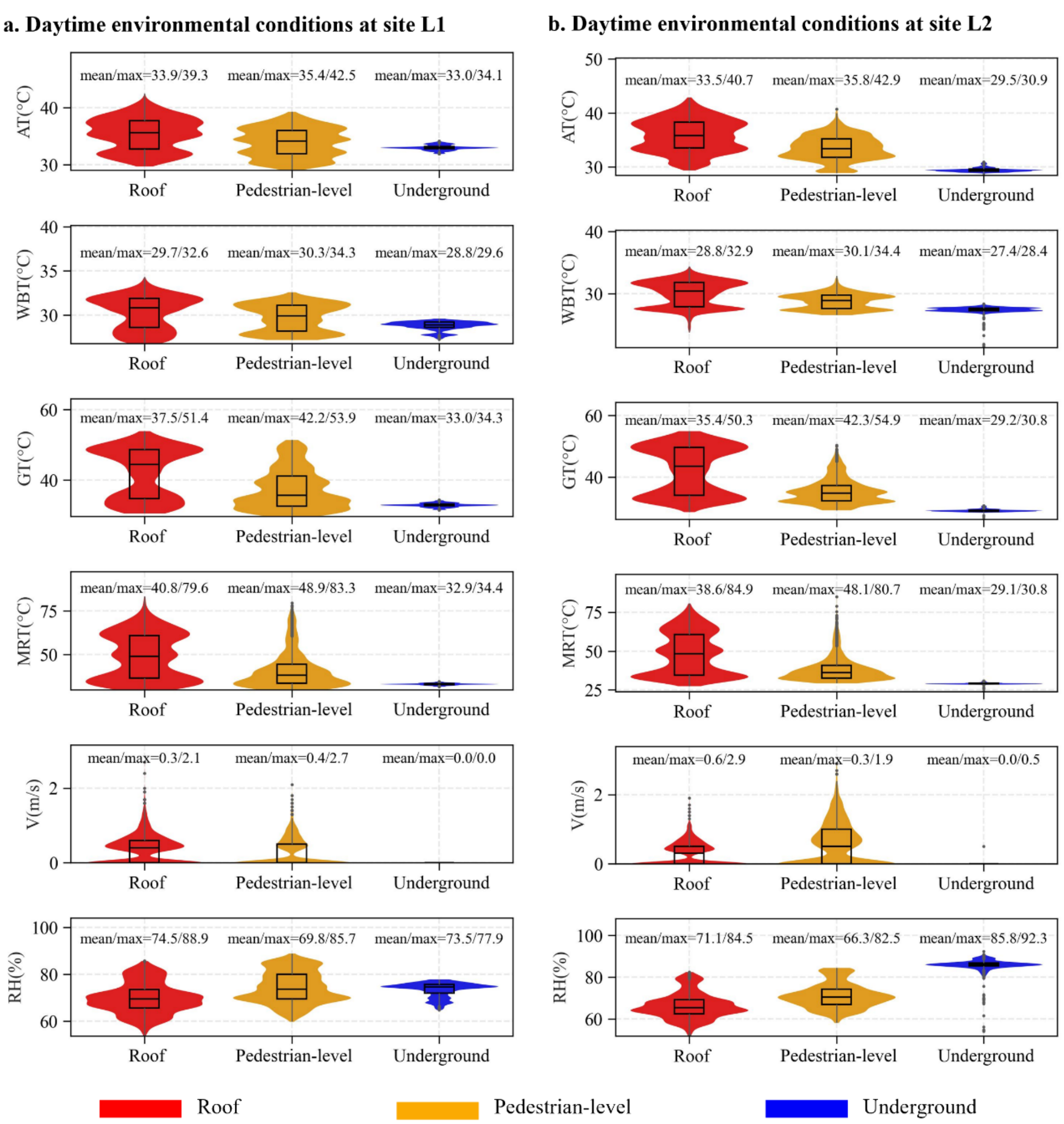

Figure 4: Environmental conditions during daytime: (a) site L1; (b) site L2

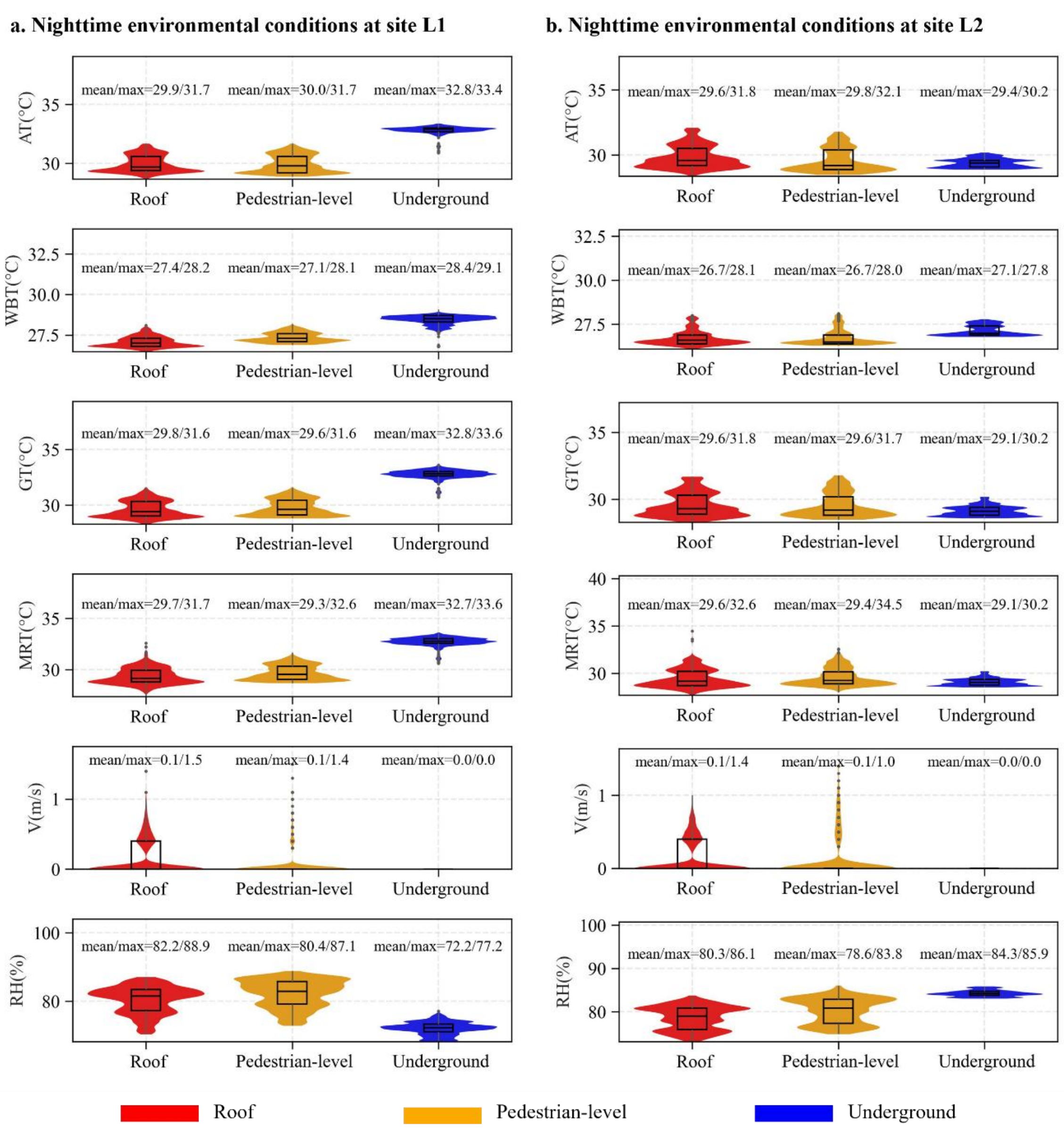

Figure 5: Environmental conditions during nighttime: (a) site L1; (b) site L2

### 3.1.1 Air temperature (AT)

During the daytime, underground AT exhibited dramatic reductions compared to aboveground environments. At Site L1, the underground had a mean AT of 33.0 °C and a maximum of 34.1 °C, while Site L2 recorded a cooler mean of 29.5 °C and a maximum of 30.9 °C. Compared to roof level, the mean AT decreased by 2.4 °C/13.4 °C (L1/L2), with corresponding reductions in maximum values of 8.4 °C/12.0 °C (L1/L2). Relative to the pedestrian-level, underground AT was lower by 0.9 °C/4.0 °C (L1/L2) in mean, and by 5.2 °C/9.8 °C (L1/L2) in maximum.

At night, underground L1 had the highest mean (32.8 °C) and maximum (33.4 °C) AT among all locations indicating poor nocturnal cooling and substantial heat retention. In contrast, underground L2 was the coolest overall (mean= 28.4 °C, max= 29.1 °C), even cooler than roof and pedestrian-level. Roof and pedestrian-level at both sites were similar with small differences (mean= 29.6–30.0 °C).

### 3.1.2 Wet-bulb temperature (WBT)

WBT values were consistently lowest in the underground environment during the daytime. Site L1 showed a mean WBT of 28.8 °C and a maximum of 29.6 °C, while Site L2 recorded a mean of 27.4 °C and a maximum of 28.4 °C. Compared to roof-level conditions, the underground had WBT reductions of 1.5 °C/2.7 °C (L1/L2) in mean and 4.7 °C/7.0 °C (L1/L2) in maximum, respectively. At pedestrian level, WBT exceeded underground values by 0.9 °C/1.4 °C (L1/L2) in mean and 3.0 °C/4.5 °C (L1/L2) in maximum, respectively.

At night, WBT followed a similar trend as AT. Underground L1 had the highest values (mean= 29.6 °C, max= 31.6 °C). Underground L2 had lower WBT (mean= 28.4 °C) but still higher than roof and pedestrian-level. The outdoor WBT ranged from 26.7 °C to 27.4 °C in mean indicating good nighttime cooling and ventilation.

### 3.1.3 Globe temperature (GT)

GT in the underground spaces demonstrated substantial reductions due to limited solar exposure during the daytime. At Site L1, GT had a mean of 33.0 °C and a maximum of 34.3 °C. At Site L2, the mean and maximum were 29.2 °C and 30.8 °C, respectively. Compared to roof-level measurements, underground represented decreases of 9.2 °C/13.1 °C (L1/L2) in mean, and 19.9 °C/24.1 °C (L1/L2) in maximum, respectively. Pedestrian-level GT surpassed the underground readings by 4.5 °C/6.2 °C (L1/L2) in mean and 17.1 °C/19.5 °C (L1/L2) in maximum, respectively.

GT was also consistently higher at night, especially at underground L1 (mean= 32.8 °C) and L2 (mean= 32.7 °C) suggesting limited radiative heat exchange. Roof and pedestrian-level recorded much lower GT (mean= 29.6–29.8 °C) indicating minimal radiant load after sunset.

### 3.1.4 Mean radiant temperature (MRT)

Underground MRT showed the largest difference across all variables indicating strong shielding from radiant heat during the daytime. At Site L1, MRT had a mean of 32.9 °C and a maximum of 34.4 °C. Site L2 showed a mean of 29.1 °C and a maximum of 30.2 °C. Compared to roof-level, the underground MRT was reduced by 16.0 °C/19.0 °C (L1/L2) in mean, and by 48.9 °C/50.6 °C (L1/L2) in maximum, respectively. Similarly, the MRT of the pedestrian-level was 7.9°C/9.5°C (L1/L2) higher than the underground environment, with the maximum MRT being 45.2°C/54.7°C (L1/L2) higher.

Compared to the daytime, the MRT at underground changed slightly at both sites (mean= 32.7 °C, max= 33.6 °C) at night indicating persistent heat from surrounding surfaces and minimal radiative cooling. As expected, the MRT dropped sharply on the roof and pedestrian-level (max= 34.5 °C).

### 3.1.5 Wind speed (V)

Air conditions were near-still in the underground environment, with a mean of 0.0 m/s at both sites during the day and night. There were reductions of 0.4 m/s (L1) and 0.3 m/s (L2) in mean, and decreases of 2.7 m/s and 1.4 m/s in maximum values compared to roof-level measurements during the daytime. Likewise, the V at the pedestrian-level exceeded those at the underground level. At night, slight air movement was detected at roof and pedestrian levels (mean= 0.1 m/s).

### 3.1.6 Relative humidity (RH)

During the daytime, RH was notably elevated at underground, with Site L1 showing a mean of 73.5% and a maximum of 77.9%, and Site L2 recording a mean of 85.8% and a maximum of 92.3%. In contrast to roof level, RH increased by 3.7% (L1) and 19.5% (L2) in mean, and by 7.8% (L1) and 9.8% (L2) in maximum. Compared with the pedestrian-level results, the RH of L1 decreased by 1.0% on average and by 11.0% at the maximum, while the RH of L2 increased by 14.7% on average and by 7.8% at the peak.

Underground L1 had the lowest nighttime RH (mean= 72.2 %, max= 77.2 %), despite having the highest temperatures indicating drier internal air or ventilation. In contrast, underground L2 had the highest RH among all locations (mean= 84.3 %) suggesting a cooler but more humid underground environment at night. Roof and pedestrian-level ranged between 78.6–82.2 % in mean.

In terms of hourly distributions (Figure 6), the period from hour 7 to 17 in the day was the period of greatest variations in temperature-related parameters (e.g., AT, WBT, GT, MRT), while humidity fluctuated significantly throughout the day. From hour 9 to 16, the AT, WBT, GT and MRT at roof and pedestrian-level began to exceed those at underground. The underground location consistently showed lower temperatures and higher humidity levels with slight variations during these hours compared to the roof and pedestrian-level locations. Wind speed exhibited less variability among the three locations.

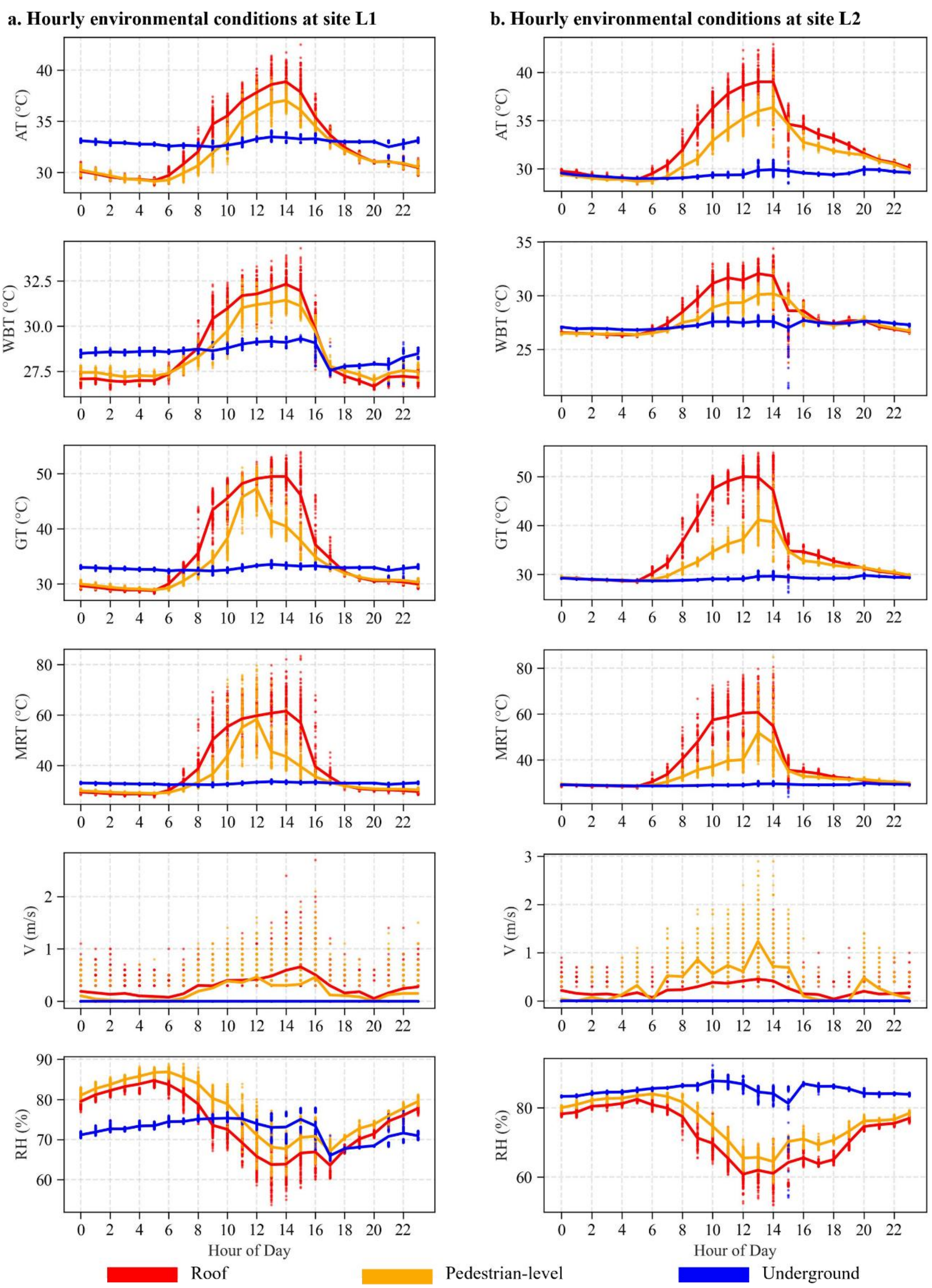

Figure 6: Hourly environmental conditions at different measurement points: (a) site L1; (b) site L2

### 3.2 Spatiotemporal variations of thermal comfort conditions

#### 3.2.1 Evaluation of decrement factor and high temperature risk

As shown in Figure 7, site L1 had a $\mu$ of 0.308, while site L2 had a lower $\mu$ of 0.197 indicating that underground structure of site L2 was more effective at reducing outdoor temperature variation over a diurnal cycle. At both sites, the underground levels maintained environmental temperature below the high temperature yellow warning threshold (<35°C) 100% of the time suggesting a consistent cooler condition during the day and night. The pedestrian-level areas stayed below the warning level more frequently than the roof level, with percentages of 81.0% for L1 and 82.5% for L2. In contrast, the roof level exceeded the high temperature threshold more often, with only 62.4%-68.4% of the time below 35°C.

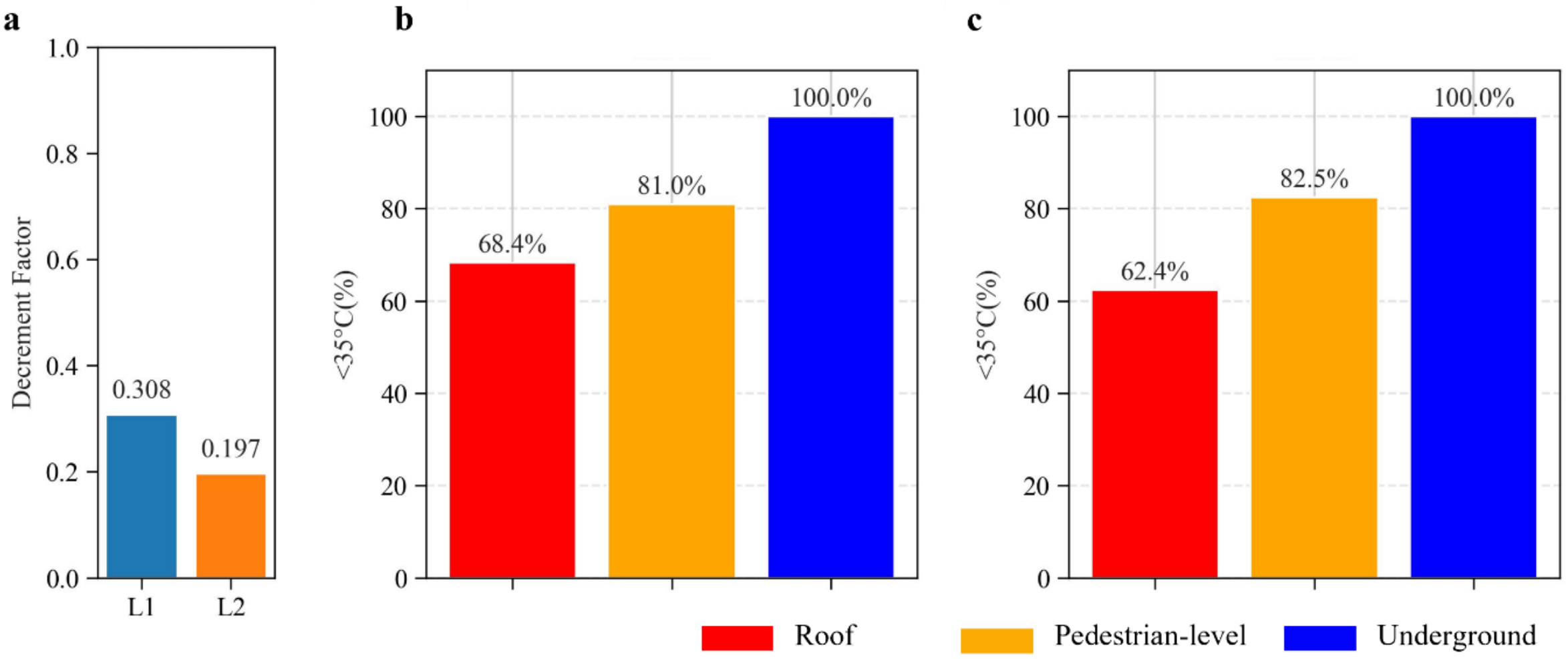

Figure 7: Evaluation of decrement factor and high temperature risk: (a) decrement factor of underground structures; (b) high temperature risk at site L1; (c) high temperature risk at site L2

#### 3.2.2 Residual analysis of sampling frequencies

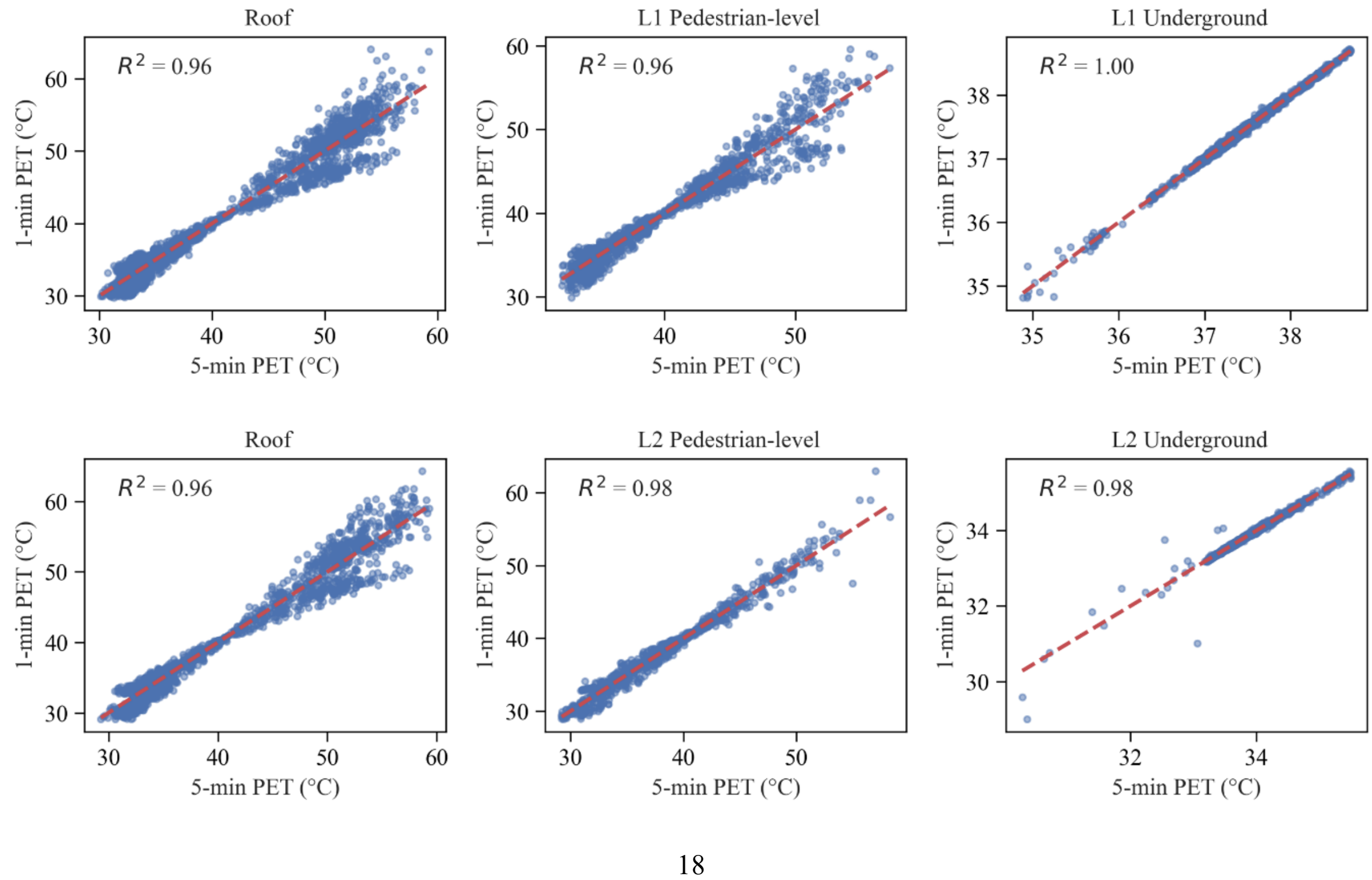

Figure 8: Residual analysis of sampling frequencies

As shown in the Figure 8, residual analysis across all points confirmed that a 5-minute sampling frequency adequately captures PET variability without compromising fidelity, and the regression models were free from resampling error. $R^2$ for the rooftop and pedestrian-level ranging from 0.96 to 0.98 indicating that at least 96 % of the variance in 1-min dataset can be explained by the 5-min dataset. For the underground level with stable environmental conditions, the residuals were minimal, with $R^2$ of 1.0 and 0.98 for site L1 and L2, respectively, suggesting a 5-min sampling interval was sufficient for interpolation.

**3.2.3 Hourly distribution of thermal comfort indices**

From the perspective of diurnal variation, as shown in Appendix B, site L1 underground space recorded maximum PET, SET, and WBGT values of 38.7 °C, 36.7 °C, and 34.1 °C, respectively, compared to 35.6 °C, 34.1 °C, and 30.8 °C at site L2. In contrast, the maximum PET values were much higher at the roof (64.0–64.3 °C) and pedestrian-level (59.6–62.9 °C). Similarly, the maximum WBGT values reached 48.5–48.6 °C at the roof and 44.6–47.3 °C at the pedestrian-level. In addition, data dispersion analysis showed that the CV of PET at the roof was 0.205–0.209, while the CV of PET at the underground space was 0.013–0.015, indicating naturally ventilated underground space not only reduced the average thermal stress but also reduced its temporal fluctuation.

Figure 9 illustrates the hourly PET distributions at different measurement points. The mean PET values were consistently higher at site L1 compared to site L2 indicating more severe thermal comfort conditions. At the pedestrian level, site L1 (37.8 °C) recorded an increase of mean and maximum PET of 1.3 °C and 4.5 °C than site L2 (36.5 °C), respectively, along with a broader range of values (L1: 33.5–49.9 °C; L2: 31.3–45.4 °C). In the underground environment, site L1 (37.5 °C) also demonstrated elevated mean PET values of 3.6°C relative to site L2 (33.9 °C), although both sites showed similar PET variance (L1: 37.1–38.2 °C; L2: 33.3–34.5 °C).

Compared to the pedestrian-level, the underground level exhibited a slight reduction in mean PET of 0.3 °C at site L1 and a more pronounced reduction of 2.6 °C at site L2. In terms of maximum hourly mean PET, the underground level demonstrated notable decreases of 11.9 °C at hour 13 at site L1 and 11.0 °C at hour 14 at site L2.

When compared to the roof level, the underground level showed a modest reduction in mean PET of 2.2 °C at site L1 and a more substantial reduction of 6.5 °C at site L2. Regarding maximum hourly mean PET, the underground level displayed significant decreases of 14.7 °C at hour 15 for site L1 and 18.3 °C at hour 14 for site L2.

In terms of temporal variations, PET at site L1 was higher in the underground space than PET at the pedestrian-level and roof from hour 18 to hour 7. The pedestrian-level PET exceeded the underground between hour 9 and hour 15, while the roof PET remained higher than the underground for a longer period (hour 8-16). At site L2, PET values across the roof, pedestrian level, and underground were similar during the early morning and nighttime hours (hour 20-6). The pedestrian-level PET began to surpass the underground from hour 10 peaking around hour 13–14. The roof PET started to exceed the underground slightly earlier from hour 7. Afterward, PET values at both the pedestrian-level and roof gradually declined and stabilized by hour 20.

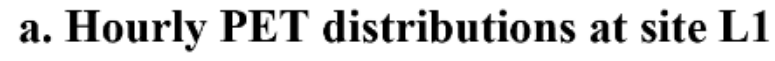

**a. Hourly PET distributions at site L1**

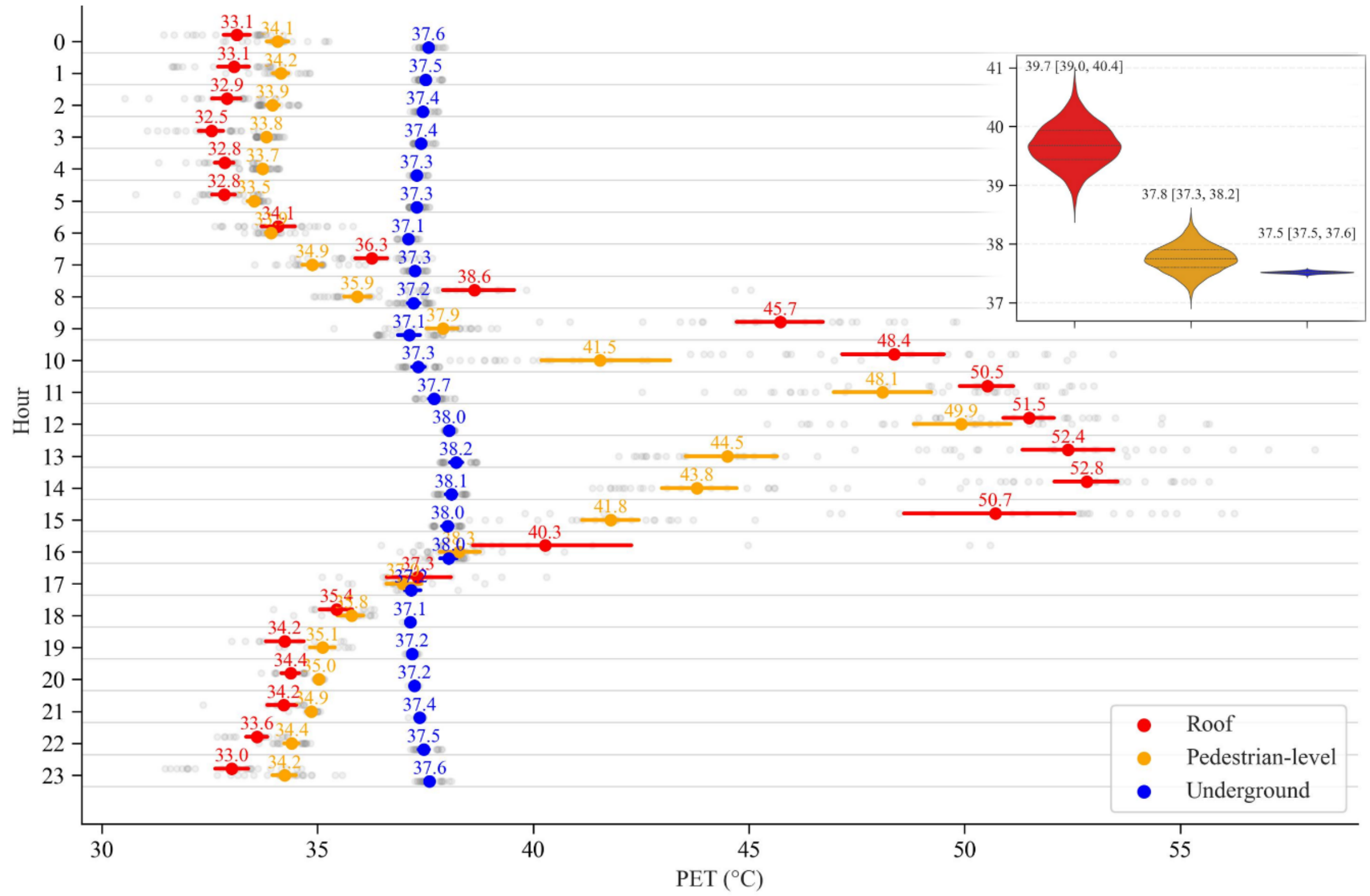

**b. Hourly PET distributions at site L2**

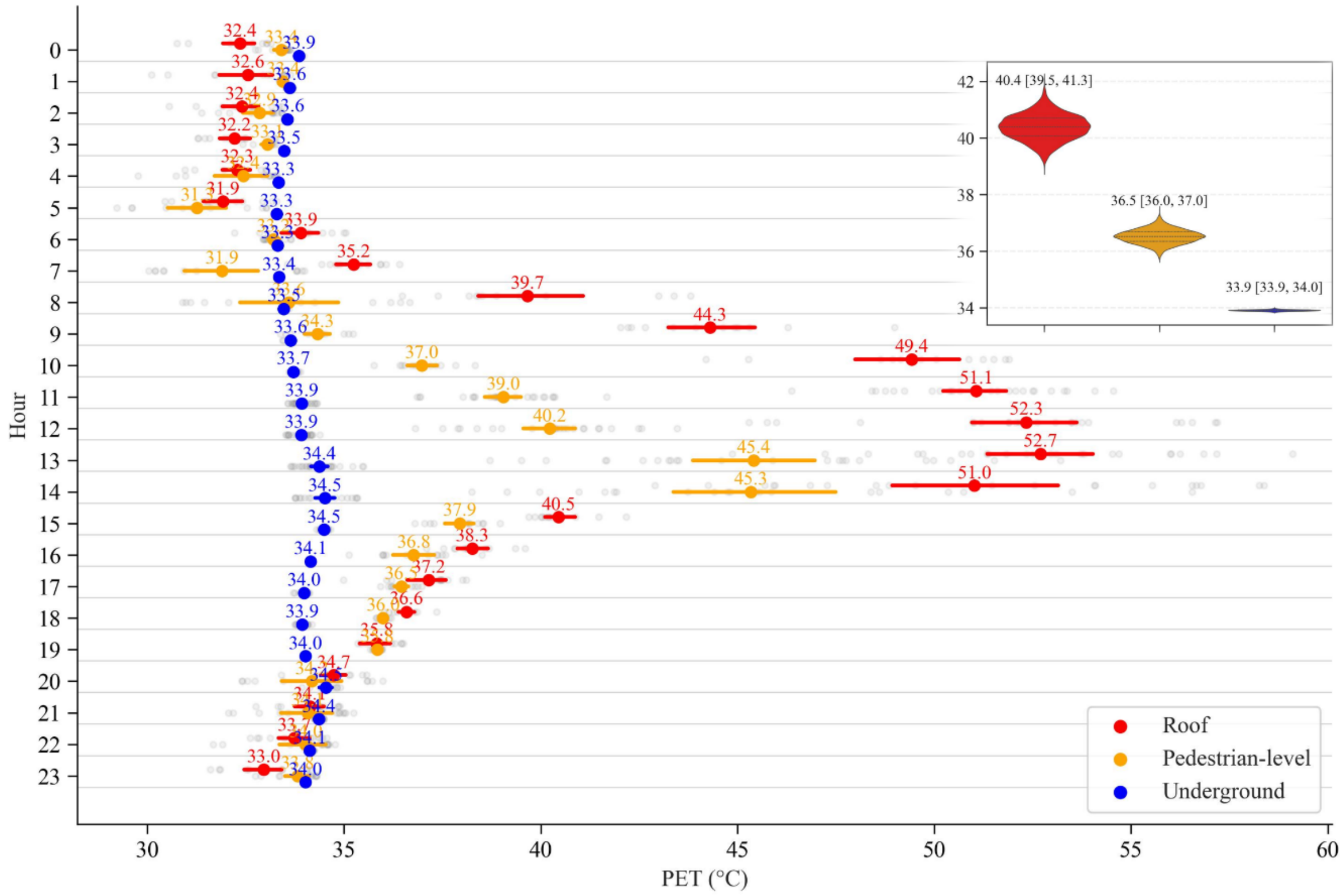

Figure 9: Hourly PET distributions at different measurement points: (a) site L1; (b) site L2

Figure 10 presents the hourly distribution of heat stress levels across different locations. At both sites, the frequency of extreme heat stress approached 100% at the roof level during the midday period (hour 9-15). Extreme thermal stresses were also observed at the pedestrian level, but less frequently than at the roof level. At site L1, pedestrian-level heat stress peaked at 100% between hour 11 and hour 14. In contrast, the pedestrian-level at site L2 experienced significantly lower exposure to extreme heat stress, ranging from 4% to 88% over the same period. The underground areas at both sites demonstrated protection against extreme heat stress. At site L1, strong heat stress persisted during both daytime and nighttime, whereas the underground environment at site L2 experienced only moderate heat stress during the corresponding periods.

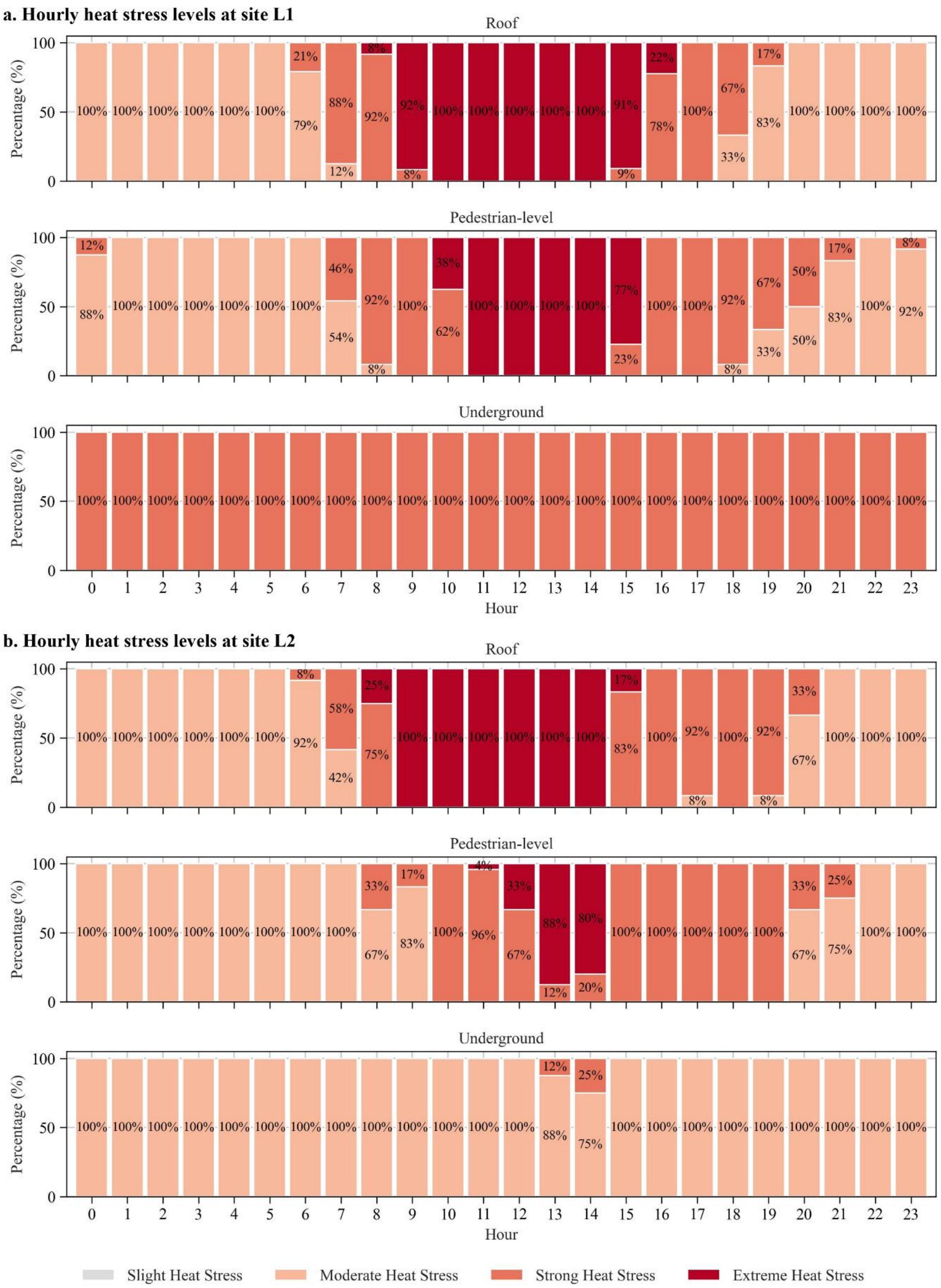

Figure 10: Hourly heat stress levels at different measurement points: (a) site L1; (b)site L2

### 3.3 Global sensitivity analysis and correlation analysis of PET

The sensitivity analysis of PET across different measurement points is presented in Figure 11. At the roof and pedestrian-level, MRT was identified as the most influential variable. Specifically, at the roof level, MRT showed S1/ST values of 0.64/0.68 for site L1 and 0.59/0.62 for site L2, respectively. At the pedestrian-level, MRT exhibited the highest sensitivity indices (L1: S1/ST= 0.72/0.73; L2: S1/ST= 0.72/0.74).

AT also showed notable contribution to roof and pedestrian-level. At the roof level, AT has S1/ST values of 0.21/0.24 for site L1 and 0.26/0.29 for site L2, respectively. Meanwhile, it was slightly lower at the pedestrian level (L1: S1/ST= 0.13/0.15; L2: S1/ST= 0.15/0.17).

Other variables (RH, V, M) demonstrated minimal effects on PET (S1/ST < 0.1). Among them, MET had a slightly greater impact than V at both roof and pedestrian-levels with ST values ranging from 0.07 to 0.09. In comparison, ST values were 0.02 to 0.04 for V. RH consistently showed very limited influence, with ST values between 0.01 and 0.05 across all locations. Changes in typical summer clothing insulation ($I_{cl}$: 0.36 to 0.57 clo) had no noticeable effect on PET.

In contrast, M emerged as the dominant factor at the underground level, with S1/ST values of 0.65 for site L1 and 0.63/0.64 for site L2. It was followed by RH (L1: S1/ST = 0.18/0.20; L2: S1/ST= 0.14/0.15), V (L1: S1/ST= 0.08/0.10; L2: S1/ST = 0.17/0.18), AT (L1: S1/ST= 0.05/0.06; L2: S1/ST= 0.04/0.04), MRT (S1/ST= 0.01), and. $I_{cl}$ showed negligible influence on PET.

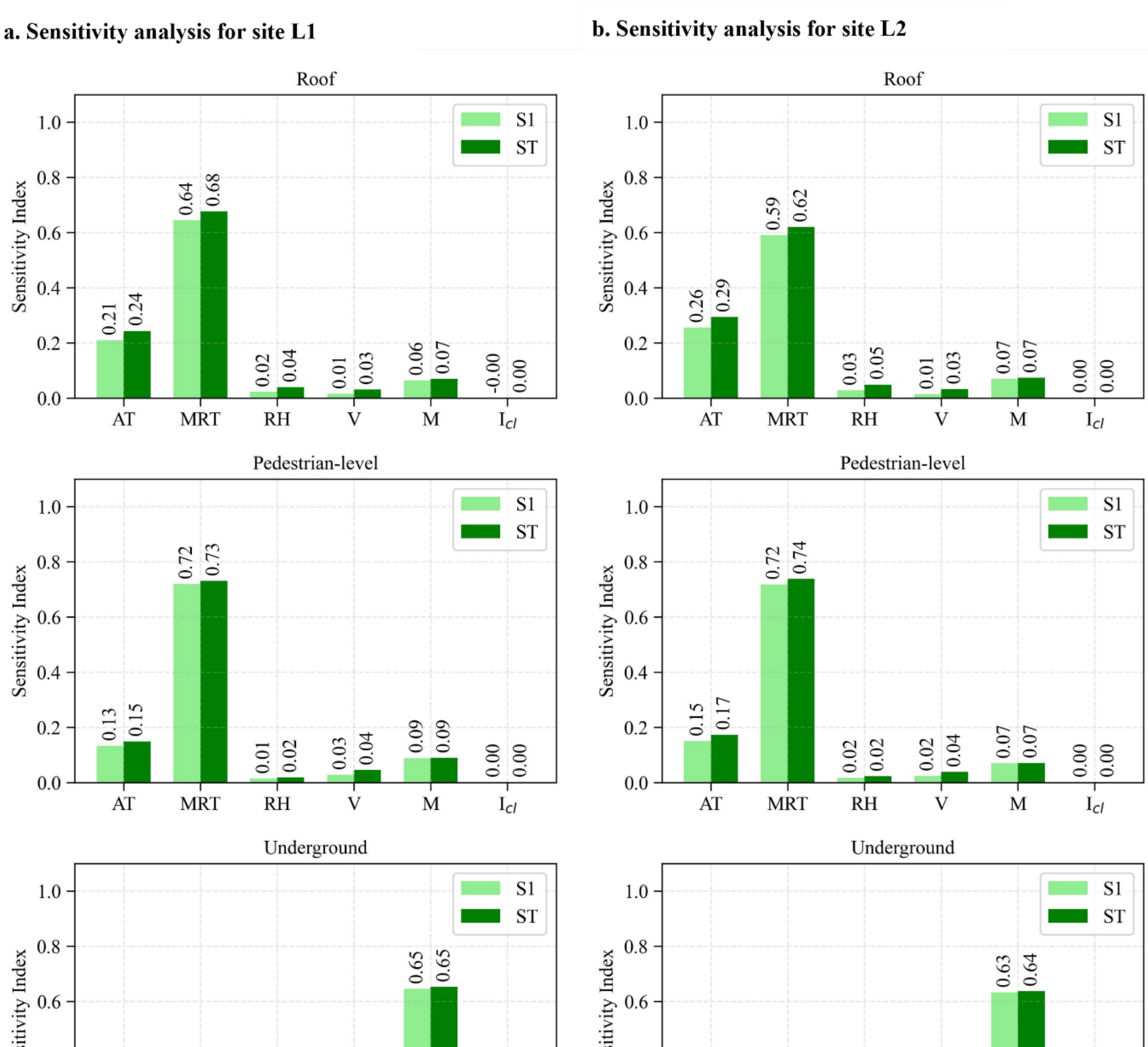

Figure 11: Sensitivity analysis of PET at different locations: (a) site L1; (b) site L2

The optimized values of hyperparameters are showing in Table 5. The regression results (Table 6) indicate strong model fit to the analytical model for both sites at underground level, with RF and XGB outperforming MLR by achieving cross-validated RMSE and SMAPE values under 0.05 °C and 0.05%, respectively. In contrast, MLR had higher error metrics with an RMSE of 0.4-0.51 °C and SMAPE of 0.84-1.22% respectively. XGB was then selected for inference of partial dependence. Figure 12 shows the partial dependences of underground PET at both sites. The results revealed that AT, MRT, RH and M were positively correlated with PET in linear form. For every 1 °C rise in AT, mean PET increased by approximately 0.5 °C to 1.0 °C. A similar trend was observed for MRT, where a 1 °C increase led to about a 0.5 °C rise in mean PET. In terms of RH, a 5 % increase resulted in roughly a 0.25 °C elevation in mean PET. M had the most pronounced effect with each 1 met increase leading to a 1.6 °C increase in mean PET. V showed a negative logarithmic effect on PET indicating that higher wind speeds were associated with lower PET values, though the cooling effect diminished at higher wind speeds. At site L1, an increase in V of approximately 1 m/s led to a 1.5 °C reduction in mean PET. In contrast, at site

L2, a similar V increase resulted in a greater cooling effect decreasing mean PET by about 2.2 °C. The coupling effect of M and RH on PET was moderate showing that higher humidity levels slightly amplified PET under elevated metabolic rates, especially at site L1 with higher AT.

Table 5: Optimized values of hyperparameters

| Model | Hyperparameter | Optimal value (Dataset L1) | Optimal value (Dataset L2) |
|---|---|---|---|
| RF | n_estimators | 200 | 200 |
| | max_depth | 20 | 20 |
| | min_samples_split | 2 | 2 |
| XGB | n_estimators | 200 | 200 |
| | max_depth | 15 | 15 |
| | learning_rate | 0.120 | 0.057 |

Table 6: Cross-validated RMSE and SAMPE of regression models

| Model | RMSE (Dataset L1) | SMAPE (Dataset L1) | RMSE (Dataset L2) | SMAPE (Dataset L2) |
|---|---|---|---|---|
| MLR | 0.4 °C | 0.84% | 0.51 °C | 1.22% |
| RF | 0.02 °C | 0.01% | 0.02 °C | 0.02% |
| XGB | 0.01 °C | 0.01% | 0.01 °C | 0.01% |

**a. Partial dependence for underground L1**

**b. Partial dependence for underground L2**

Figure 12: The partial dependence analysis of PET at the underground level: (a) site L1; (b) site L2

As shown in the one-way ANOVA summary evaluating the effect of M on PET across different locations (Table B8, Appendix B), PET was significantly influenced by M at all locations (p-value < 0.05). The F-statistic increased significantly from roof to underground, supporting previous sensitivity analysis that underground environments were more strongly regulated by internal factors such as metabolism. The distribution of PET values at various metabolic rates ranging from 1 to 5 met at different locations was shown in the Figure 13. On the roof, PET values gradually increased with metabolic rate reaching the highest PET of 47.5°C at 5 met. At the pedestrian-level, PET values also increased with metabolic rate peaking at 44.4°C at 5 met. In contrast, the underground location exhibited the lowest PET values across all metabolic rates, with the highest PET being 42.5°C at 5 met corresponding to activity level of approximately 2.6 met and 3.8 met at roof and pedestrian-level location, respectively. This result was consistent with previous findings that underground locations provided a cooler environment compared to rooftop and pedestrian locations, especially at higher metabolic rates.

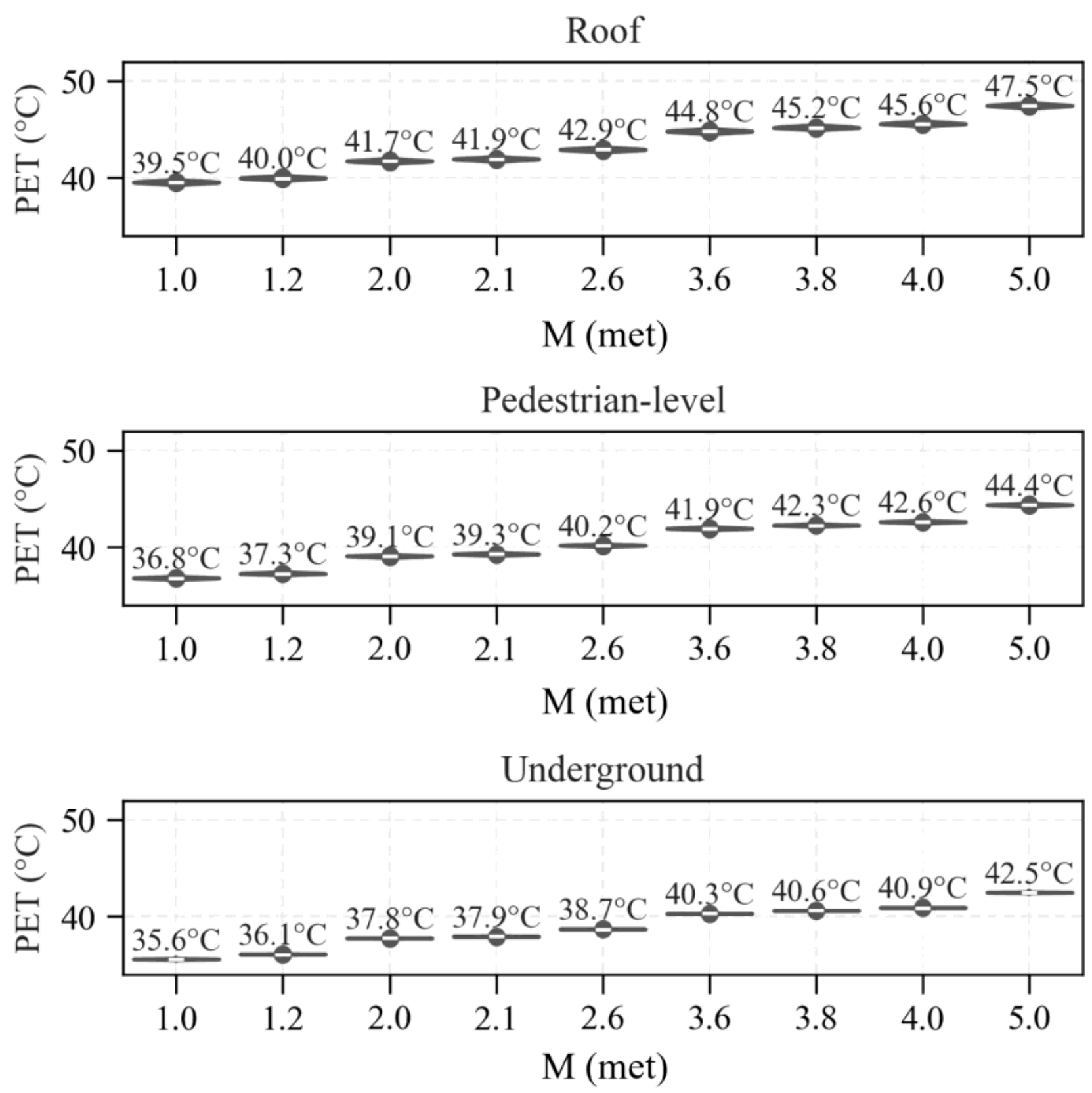

Figure 13: The impact of human metabolism on PET at different locations

## 4. Discussions

### 4.1 Applicability of thermal comfort indices in underground spaces

Underground spaces cover a wide range from deep facilities (e.g., deep mines) to shallow spaces (e.g., urban parking lots), each with distinct thermal characteristics. In this study, the roof, pedestrian-level space, and underground space each present a unique thermal environment, and thermal comfort indices must be able to account for differences in radiation, humidity, wind, and air temperature, as well as individual differences such as clothing and metabolic rates.

Some limitations of thermal comfort indices were observed. The valid input ranges for UTCI and Out_SET* are limited (e.g., UTCI: -50 < AT (°C) < 50, AT-70 <MRT (°C) < AT+30, 0.5 m/s < V (m/s) < 17.0; Out_SET*: 10 < AT (°C) < 40, 10 < MRT (°C) < 40, 0 < V (m/s) < 2, 1 < M (met) < 4, and 0 < $I_{cl}$ (clo) < 1.5) (Tartarini and Schiavon, 2020), leading to limited output for scenarios with high MRT and low V (Appendix B, Figure B2). While WBGT provided a continuous data result (Appendix B), its value depends only on WBT, GT, and AT, without accounting for convective heat loss (wind speed), human metabolism, and personal clothing.

In contrast, PET calculation incorporates the radiation environment, metabolic rate, and clothing insulation, making it more suitable for comparing thermal comfort in different urban environments. It is an important index for evaluating human thermal comfort in outdoor environments, particularly in the context of urban heat islands and occupational health and safety (Farhadi et al., 2019; Fiorillo et al., 2023; Hwang et al., 2023). Derived from the Munich Energy Balance Model for Individuals (MEMI) by Hoppe, PET represents the air temperature in a wind and solar radiation free indoor environment yielding the same core and skin temperature as the actual outdoor conditions (Höppe, 1999). Walther and Goestchel propose a correction to the PET calculation defined by Hoppe and compare the corrected model with the original method, highlighting errors in calculation routine and vapor diffusion model (Walther and Goestchel, 2018). Given its ability to account for complex outdoor conditions (e.g., solar exposure and ventilation), PET is well-suited for evaluating transitional environments such as underground entrances, semi-outdoor spaces, and shaded pedestrian areas.

### 4.2 Potential and challenges of underground space as heat shelter

#### 4.2.1 Reduced AT and MRT in underground spaces

Overall, the underground space exhibited distinct environmental conditions (e.g., AT, WBT, GT, MRT, RH, V) compared to roof and pedestrian-level locations (p-value of KS test <0.01). Particularly, compared to pedestrian-level, the underground space exhibited the lowest AT and MRT reducing maximum AT by 5.2 °C-9.8 °C maximum MRT by 45.2°C-54.7°C during the day suggesting a relatively cooler environment. The increasing PET from underground to the rooftop indicated a vertical temperature gradient in urban spaces highlighting the potential of underground space serving as the heat shelter. The cooling effect of underground space has been reported by few studies. Field measurements and simulations showed that underground rooms had 3 °C lower indoor temperatures in summer due to reduced heat transfer, with application potential for improved thermal comfort and energy efficiency in hot climates (Alwetaishi et al., 2021). Another study found that urban underground parking can effectively mitigate urban heat islands and improve outdoor thermal comfort. Simulations showed urban underground parking lowered peak air temperature by 1.5 °C and mean temperature by 0.67 °C, and reduced mean radiant temperature from 55–65 °C to 20–27 °C after development (Yang et al., 2019). Even the sunken plaza near the entrance of the underground metro station exhibited reduced air temperature (Zhao et al., 2023).

#### 4.2.2 Reduced V and elevated RH in underground spaces

However, the absence of natural ventilation (0 m/s) observed in this study indicates that maintaining thermal comfort in underground spaces is challenging. While underground areas may offer a consistently cooler environment, they can also experience the highest humidity levels within the diurnal cycle as observed at site L2. The combination of high humidity and poor ventilation can impose significant thermal stress on occupants despite the lower air temperatures. High humidity can significantly increase perceived temperature reducing the ability of thermoregulation through evaporation of sweat. In the absence of ventilation, this effect is compounded as it prevents heat and moisture releases. As a result, underground areas could still feel hot and stuffy due to high humidity and poor ventilation. These issues were common in underground space. A study on underground refuge chambers found that high air temperature, humidity, and $CO_2$ levels significantly impacted thermal sensation and acceptability, with TSV rising significantly at elevated $CO_2$, especially at RH of 85% (Li et al., 2018b). Temperature and RH significantly impact thermal responses, with thermal comfort decreasing as temperature and RH increase in deep underground spaces (Yang et al., 2022). Similarly, a metro station study found that high temperatures and low wind speeds significantly increased occupant discomfort emphasizing the need for region-specific environmental controls to improve comfort and energy efficiency in subtropical metro systems (Fang et al., 2025). Nevertheless, it is very interesting and counterintuitive that a low temperature (mean AT=28.4 °C) and high humidity (mean RH= 84.3%) combination like site L2 (PET: 33.3–34.5 °C, moderate heat stress) resulted in a better heat stress level than site L1 (PET: 37.1–38.2 °C, strong heat stress) with high temperature (mean AT= 32.8 °C) and low humidity (mean RH= 72.2%), despite poor ventilation at both sites. This is because under high temperature conditions, a high humidity environment significantly inhibits human evaporative cooling, thereby exacerbating heat stress. At relatively low temperature, the effect of high humidity on thermal discomfort is weakened leading to lower PET level.

**4.2.3 Higher tolerance for intensive physical exercises in underground spaces**

At the underground level, M was the most influential factor (S1/ST= 0.63-65). Higher metabolic heat in enclosed spaces significantly increases thermal discomfort, with activity change of 1 met resulting in an approximately 1.6 °C increase in PET. Sun shading (Chen et al., 2020; Chen et al., 2023; de Abreu-Harbich et al., 2015; Donovan et al., 2021; Elgheznawy and Eltarabily, 2021; Li et al., 2021; Ou and Lin, 2023; Yu et al., 2020) and high thermal mass (Caliot et al., 2024; Chen et al., 2023) are well-known for their cooling benefits. Compared with sun-exposed areas, the naturally ventilated underground space offers a thermally safer microenvironment where short-duration and high-intensity activities remain physiologically feasible.

It is important to note that higher humidity slightly amplified PET under elevated metabolic rates as shown in the partial dependence results. Nevertheless, a related study discovered that higher humidity had little effect on reducing overall dissatisfaction of at M higher than 1.6 met for sedentary subjects, but there were noticeable differences in non-sedentary activities (Fountain et al., 1999).

**4.2.4 Influences of aboveground morphologies on underground thermal comfort**

As indicated by the lower CV values, the underground space exhibited more stable environmental conditions and thermal comfort levels than the aboveground spaces. This stability can be attributed to the high thermal inertia and substantial thermal mass of the surrounding soil and concrete structure, moderating short-term fluctuations in air temperature and solar radiant load. Although both sites achieved similar reductions in PET between pedestrian and underground levels (L1: 11.9 °C; L2: 11.0 °C), the consistently higher PET values at site L1 (pedestrian-level: ΔT= 4.5 °C; underground: ΔT= 3.6 °C) and lower $\mu$ (0.197) suggested that the elevated underground PET at site L1 may result from a combination of higher ground surface temperatures and less effective thermal dampening compared to site L2. The higher ground surface temperature could be attributed to reduced greenery and limited shaded areas, which could be mitigated by planting high Leaf Area Index (LAI) species (Morakinyo et al., 2017) and installing outdoor shading devices to reduce the sky view factor (Deng et al., 2023; Yang et al., 2010; Yu et al., 2025; Yu et al., 2020).

**4.2.5 Potential of ventilation cooling**

This study highlights the need to balance airflow, temperature and humidity control in underground spaces and identified V as significant contributing factor to the PET. A negative logarithmic relationship was observed between V and PET, where an increase of 1 m/s reduced mean PET by approximately 1.5–2.2 °C. However, the cooling effect diminished significantly beyond 1.5 m/s. Ventilation cooling is also crucial for thermal comfort of miners in high geothermal tunnels (Zhou et al., 2023). A study conducted in hot and humid underground mining environments used sensitivity analysis to assess the impact of environmental factors and metabolic rate on thermal comfort (Sunkpal et al., 2018). It found that an air velocity of 1.5 m/s provided optimal thermal comfort, while higher metabolic rates increased heat strain requiring lower ambient temperatures to sustain effective sweat evaporation and maintain comfort.

Furthermore, the cooling effect of ventilation can be maximized by considering transient thermal comfort when moving between hotter and cooler spaces. In this study, underground areas offered a cooler period from hour 10 to hour 16, during which additional cooling could greatly enhance comfort. Previous study indicates that the combination of rapid temperature drops and enhanced air movement leads to higher thermal comfort with rapid stabilization (Yao et al., 2019).

**4.3 Integrated aboveground and underground design**

This study proposes a design strategy for an integrated aboveground and underground environment (Figure 14). Exposed outdoor areas are most vulnerable to extreme heat and require urgent adaptative measures (e.g., radiant cooling, green roofs, shading). At the pedestrian level, air temperature is primarily influenced by the sky view factor representing the amount of solar radiation received. (Yu et al., 2025). Therefore, implementing street shading solutions (e.g., trees, parasols) can help reduce direct sunlight and mitigate heat stress. Previous studies have shown that greenery have a significant cooling effect on both daytime and nighttime thermal environments (Ouyang et al., 2020; Wong et al., 2021). For example, green facades with a LAI of 4.8 can reduce surface temperatures by 3.6°C to 4°C in summer and by 1.8°C in spring with an LAI of 3.6 (Pérez et al., 2022). Another study reported that trees with LAI of 4.0 and a canopy radius of 5.0 m were optimal for street aspect ratio of 1.0–3.0, which can provide a maximum UTCI cooling of approximately 4.0 °C (Chen et al., 2023). Field measurements also confirmed that parasols had a cooling effect by reducing the mean UTCI from 45.7 °C to 39 °C (Zhang et al., 2023). Commercial stores located underground can benefit from the high thermal mass of the aboveground structure. The exposed curtain wall is enhanced with low-e glass while remaining transparent to take advantage of daylight. However, the nighttime heat retention and poor ventilation highlight the need for improved underground ventilation strategies. Previous studies have shown that integrated greenery and natural ventilation are effective measures to improve the underground environment (Wen et al., 2024a, 2024b, 2023). Therefore, the underground open space is designed as a semi-outdoor space planted with greenery and introduced with natural ventilation to support intensive activity space (e.g., children's playground, basketball court, running track).

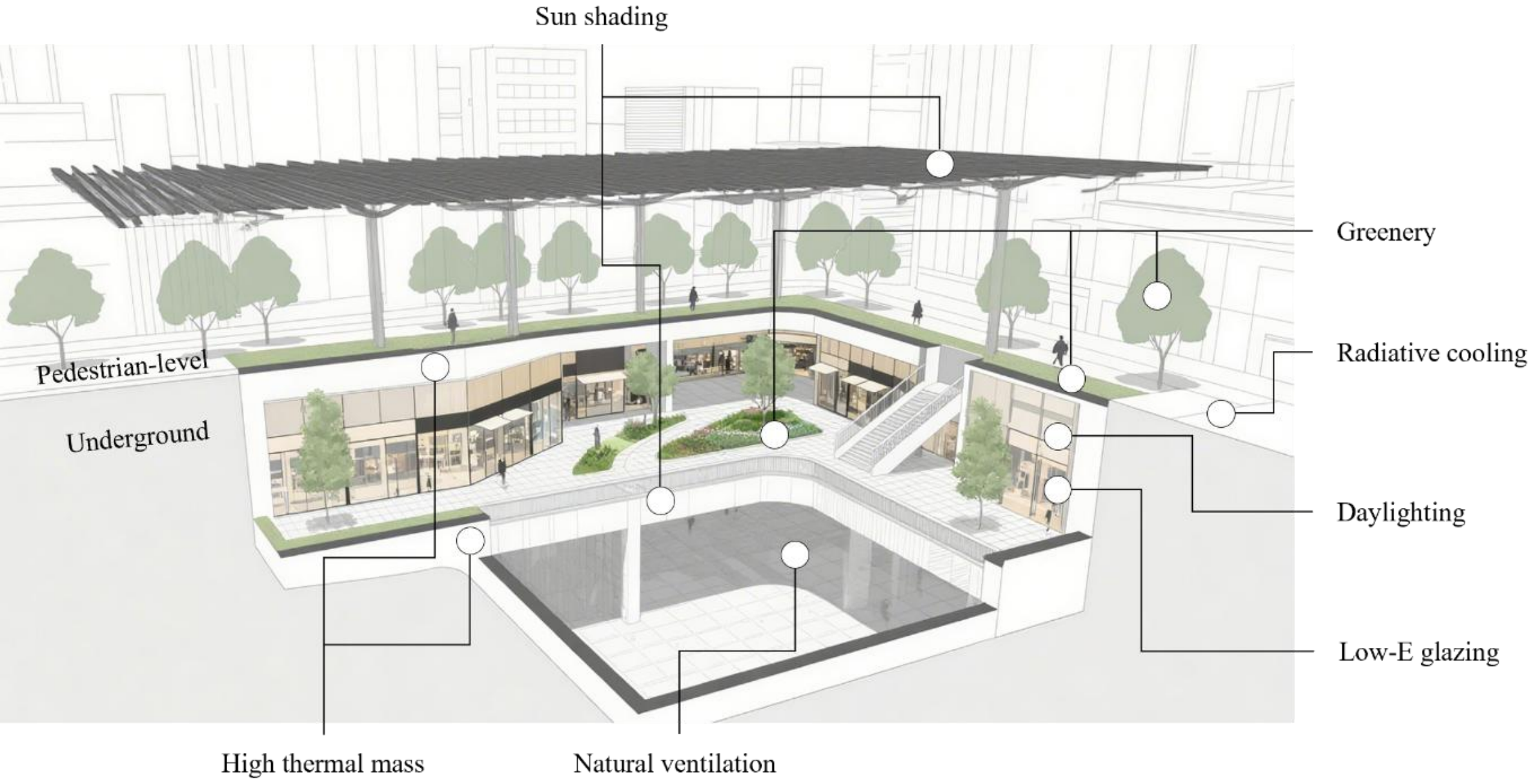

Figure 14: Design strategies for sustainable and comfortable underground environments

**4.4 Limitations and future studies**

In terms of limitations, this study used a naturally ventilated underground parking lots as the investigation site and focused specifically on high temperature conditions based on short-term datasets for analysis, which may not fully represent other underground environments (e.g., subways) or climate conditions (e.g., cool and cold conditions). Therefore, it is recommended to include seasonal experiments to enable cross-comparisons under different microclimate conditions. The integrated radiation method can further improve the estimation of mean radiation temperature based on local climatic conditions.

The PET calculation relied on static metabolic rate and clothing insulation assuming lightweight summer clothing, which may lead to differences between the sample and the population. Future studies shall conduct site analysis of different underground environments and apply more detailed human metabolic rates and clothing insulation to calculate PET. In addition, future research needs to measure and compare other environmental factors (e.g., illuminance level, $CO_2$ level, sound pressure level) to provide a holistic assessment of different environments. Local calibration of mean radiant temperature based on integral radiation method using pyranometers and pyrgeometers is recommended for future research to reduce the instrument errors that may be caused by globe thermometer. This study was limited to objective microclimatic monitoring and PET simulation as an initial assessment of extreme heat mitigation in naturally ventilated underground space. Future study shall include a questionnaire survey (e.g., TSV, TCV) to validate comfort ranges through subjective responses.

To further validate the effectiveness of proposed design strategies, parametric simulations with varying key design parameters such as shading coefficient, greenery coverage, and ventilation opening ratios could be conducted. Simulation tools such as ENVI-met could then be employed to quantify the impact of these variables on thermal comfort indicators under extreme summer conditions

## 5. Conclusions

The study investigated the spatiotemporal variations of environmental conditions across roof, pedestrian-levels, and underground levels at two sites (L1 and L2). During daytime, underground areas had lower AT, MRT, and V, and higher RH compared to aboveground space. For example, mean underground AT was 33.0°C at L1 and 29.5°C at L2, with maximum values of 34.1°C and 30.9°C, respectively. Compared to roof level, mean AT decreased by 2.4°C/13.4°C (L1/L2), and maximum AT decreased by 8.4°C/12.0°C (L1/L2). Underground spaces consistently maintained temperatures below the high-temperature threshold (<35°C).

The hourly distribution of PET showed that underground PET values were consistently lower than those at roof and pedestrian-levels, especially during peak hours (hour 8-16). Both sites achieved similar PET reductions between pedestrian and underground levels (L1: 11.9°C; L2: 11.0°C). The higher pedestrian-level PET at L1 could be due to reduced greenery and limited shaded areas. The $\mu$ was 0.308 for L1 and 0.197 for L2 indicating better thermal dampening at L2.

Sensitivity analysis revealed MRT was the most influential variable at roof and pedestrian levels, with S1/ST ranging from 0.59 to 0.72. AT also had notable contributions with S1/ST values between 0.13 and 0.26. In contrast, at the underground level, M was the dominant factor with S1/ST values of 0.63-0.65, followed by RH (0.14-0.20) and V (0.08-0.18). Partial dependence analysis showed 1°C increase in AT and MRT led to approximately 0.5-1.0°C and 0.5°C increases in PET, respectively. M had the most pronounced effect with a 1 met increase leading to a 1.6°C rise in PET. V showed a negative logarithmic effect on PET with a 1 m/s increase reducing PET by 1.5-2.2°C. Underground showed higher tolerance for intensive physical activities.

In summary, underground spaces can serve as effective heat shelters but require improved ventilation and humidity control. Aboveground areas need adaptive strategies such as radiant cooling, green roofs, and shading (e.g., trees, parasols) to mitigate extreme heat. Semi-outdoor underground designs with greenery and natural ventilation can enhance thermal comfort for high-activity use. The results of this study were mainly applicable to extreme high temperature scenarios and recommended parameters were preliminary suggestions requiring further validation. Future research should combine seasonal data to verify thermal comfort conditions of underground spaces throughout the year.

**Acknowledgements**

The authors greatly appreciate the following funding support for this research: (1) The Fujian Province Young and Middle-aged Teacher Education and Research Project Funding (JAT241009); (2) Fuzhou University Research Starting Fund (511470); (3) Fuzhou University Testing Fund of Precious Apparatus (2025T031); (4) Shanghai Science and Technology Innovation Action Plan for 2022 (22dz1207100); (5) Tongji Architectural Design (Group) Co., Ltd. (2022J-JZ01, 2023J-JB04).

The authors would like to thank anonymous reviewers for their constructive comments and suggestions, which greatly improved the content of this article.

**Conflicts of Interest**: The authors declare no conflict of interest.

**Appendix A: Specifications of environmental sensors**

The Kestrel 5400 equips with a 25mm black copper globe, and its GT readings are converted internally to represent the GT measured on a standard 150mm globe (Naughton, 2020). The measurements were performed at a height of 1.5 m.

Table A1: The specifications of environmental sensors

| Environmental parameters | Range | Resolution | Accuracy |
|---|---|---|---|
| Dry-bulb air temperature | -29 to 70 °C | 0.1 °C | ±0.5 °C |
| Relative humidity | 10 to 90% | 0.1 %RH | ±2%RH |
| Wind speed | 0.6 to 40 m/s | 0.1 m/s | ±3% |
| Globe temperature | -29 to 60 °C | 0.1 °C | ±1.4 °C |

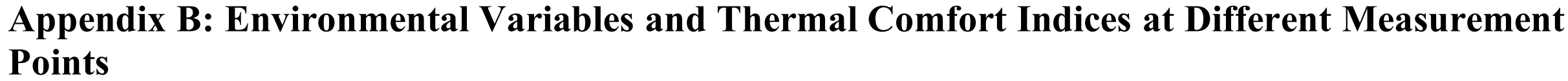

## Appendix B: Environmental Variables and Thermal Comfort Indices at Different Measurement Points

a. Environmental conditions at site L1

b. Environmental conditions at site L2

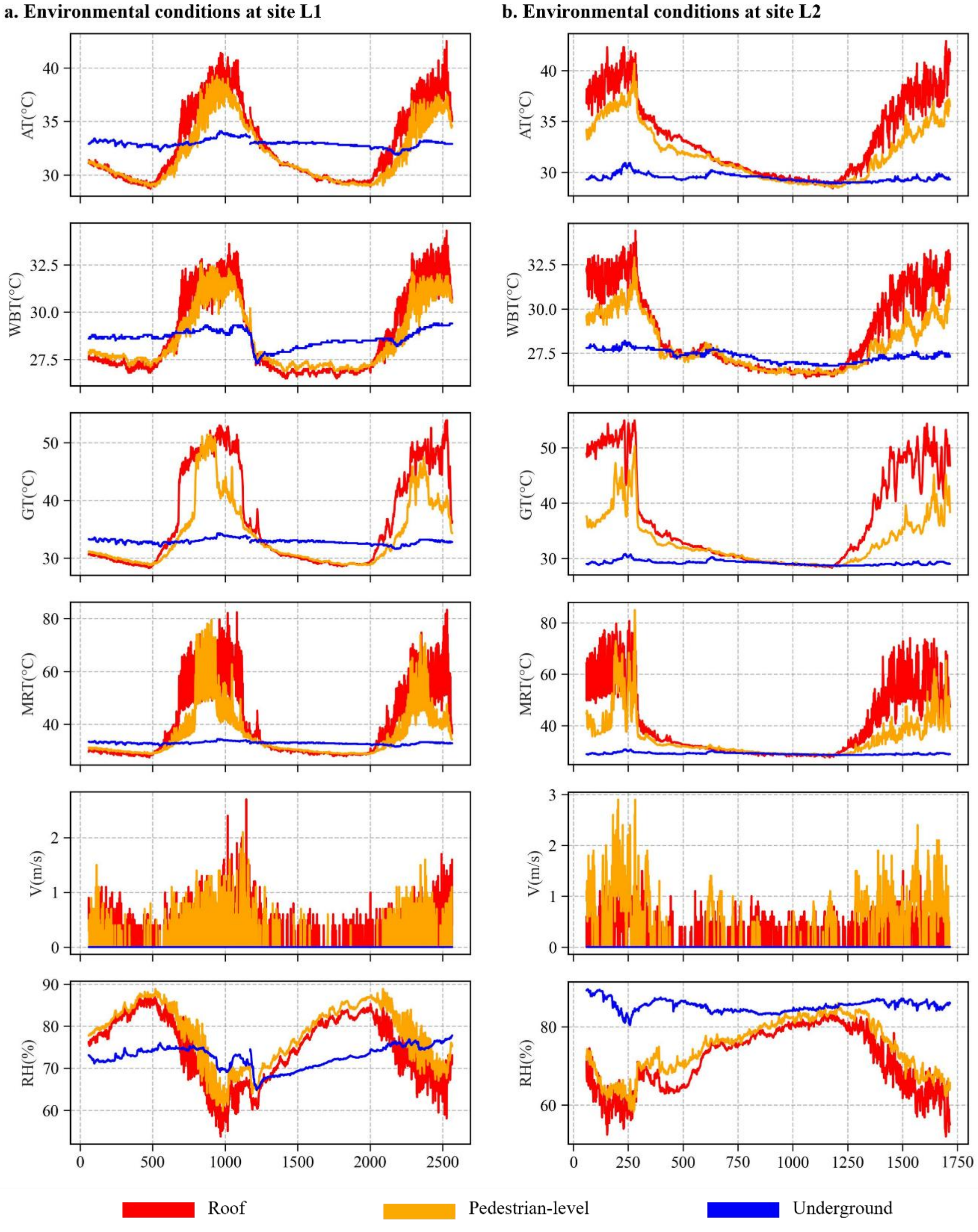

Figure B1: Time series plots of measured environment variables: (a) site L1; (b) site L2

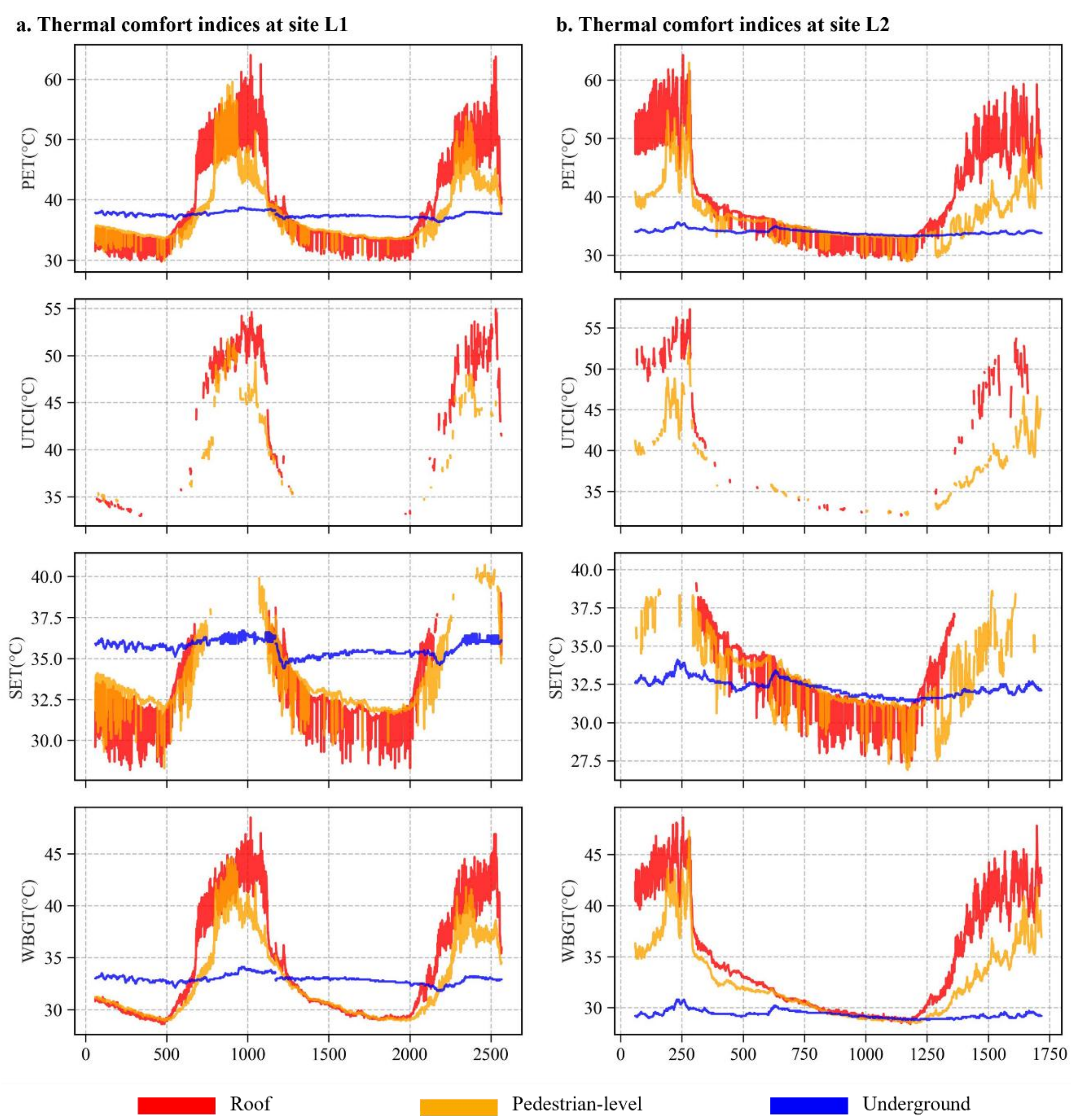

Figure B2: Time series plots of calculated thermal comfort indices at different locations: (a) site L1; (b) site L2

Table B1: Descriptive statistics of environmental variables at site L1

| Site L1 | Variable | Mean | Min | 25% | 50% | 75% | Max | Std | CV |
|---|---|---|---|---|---|---|---|---|---|
| Roof | AT(°C) | 32.9 | 28.7 | 29.8 | 31.3 | 35.8 | 42.5 | 3.5 | 0.106 |
| | GT(°C) | 36.3 | 28.3 | 29.5 | 31.7 | 45.3 | 53.9 | 8.2 | 0.226 |
| | WBT(°C) | 28.8 | 26.5 | 27.0 | 27.6 | 30.9 | 34.3 | 2.1 | 0.073 |
| | V(m/s) | 0.3 | 0.0 | 0.0 | 0.0 | 0.5 | 2.7 | 0.3 | 1.000 |
| | RH(%) | 74.8 | 53.7 | 69.0 | 76.0 | 81.7 | 87.1 | 7.6 | 0.102 |
| | MRT(°C) | 39.7 | 27.4 | 29.2 | 31.8 | 50.0 | 83.3 | 13.6 | 0.343 |
| Pedestrian-level | AT(°C) | 32.0 | 28.9 | 29.8 | 31.0 | 34.3 | 39.3 | 2.7 | 0.084 |
| | GT(°C) | 33.9 | 28.8 | 29.6 | 31.1 | 36.4 | 51.4 | 5.7 | 0.168 |
| | WBT(°C) | 28.6 | 26.9 | 27.3 | 27.8 | 30.2 | 32.6 | 1.6 | 0.056 |
| | V(m/s) | 0.2 | 0.0 | 0.0 | 0.0 | 0.4 | 2.1 | 0.3 | 1.500 |
| | RH(%) | 78.1 | 59.8 | 72.7 | 79.2 | 84.2 | 88.9 | 6.9 | 0.088 |
| | MRT(°C) | 35.6 | 28.7 | 29.6 | 31.1 | 38.9 | 79.6 | 9.4 | 0.264 |
| Underground | AT(°C) | 32.9 | 30.9 | 32.7 | 32.9 | 33.1 | 34.1 | 0.4 | 0.012 |
| | GT(°C) | 32.9 | 30.7 | 32.6 | 32.9 | 33.1 | 34.3 | 0.5 | 0.015 |
| | WBT(°C) | 28.6 | 26.8 | 28.4 | 28.6 | 28.9 | 29.6 | 0.5 | 0.017 |
| | V(m/s) | 0.0 | 0.0 | 0.0 | 0.0 | 0.0 | 0.0 | 0.0 | N/A |
| | RH(%) | 72.9 | 64.8 | 71.4 | 73.0 | 74.8 | 77.9 | 2.7 | 0.037 |
| | MRT(°C) | 32.8 | 30.6 | 32.6 | 32.9 | 33.1 | 34.4 | 0.5 | 0.015 |

Table B2: Descriptive statistics of environmental variables at site L2

| Site L2 | Variable | Mean | Min | 25% | 50% | 75% | Max | Std | CV |
|---|---|---|---|---|---|---|---|---|---|
| Roof | AT(°C) | 33.6 | 28.4 | 30.0 | 33.1 | 37.0 | 42.9 | 3.8 | 0.113 |
| | GT(°C) | 37.6 | 28.3 | 29.8 | 33.4 | 47.4 | 54.9 | 8.7 | 0.231 |
| | WBT(°C) | 28.8 | 23.8 | 26.7 | 27.8 | 31.2 | 34.4 | 2.3 | 0.080 |
| | V(m/s) | 0.3 | 0.0 | 0.0 | 0.0 | 0.5 | 1.9 | 0.3 | 1.000 |
| | RH(%) | 70.9 | 51.9 | 64.2 | 70.4 | 78.4 | 83.8 | 7.9 | 0.111 |
| | MRT(°C) | 41.2 | 27.5 | 29.6 | 33.8 | 51.3 | 80.7 | 14.1 | 0.342 |
| Pedestrian-level | AT(°C) | 32.0 | 28.5 | 29.3 | 31.6 | 34.1 | 40.7 | 2.7 | 0.084 |
| | GT(°C) | 33.2 | 28.5 | 29.6 | 31.9 | 35.5 | 50.3 | 4.3 | 0.130 |
| | WBT(°C) | 28.0 | 26.3 | 26.7 | 27.6 | 29.3 | 32.9 | 1.4 | 0.050 |
| | V(m/s) | 0.4 | 0.0 | 0.0 | 0.0 | 0.7 | 2.9 | 0.6 | 1.500 |
| | RH(%) | 74.6 | 58.2 | 69.4 | 75.3 | 80.8 | 86.1 | 6.7 | 0.090 |
| | MRT(°C) | 35.2 | 28.0 | 29.7 | 32.1 | 37.8 | 84.9 | 7.9 | 0.224 |
| Underground | AT(°C) | 29.4 | 28.5 | 29.2 | 29.4 | 29.6 | 30.9 | 0.4 | 0.014 |
| | GT(°C) | 29.2 | 26.3 | 28.9 | 29.1 | 29.4 | 30.8 | 0.4 | 0.014 |
| | WBT(°C) | 27.3 | 21.4 | 27.0 | 27.3 | 27.7 | 28.4 | 0.5 | 0.018 |
| | V(m/s) | 0.0 | 0.0 | 0.0 | 0.0 | 0.0 | 0.5 | 0.0 | N/A |
| | RH(%) | 85.2 | 54.1 | 84.2 | 85.4 | 86.4 | 92.3 | 2.6 | 0.031 |
| | MRT(°C) | 29.1 | 23.9 | 28.8 | 29.1 | 29.3 | 30.8 | 0.4 | 0.014 |

Table B3: Descriptive statistics of thermal comfort indices at site L1

| Site L1 | Variable | Mean | Min | 25% | 50% | 75% | Max | Std | CV |
|---|---|---|---|---|---|---|---|---|---|
| Roof | PET(°C) | 39.6 | 29.7 | 33.6 | 35.6 | 46.8 | 64.0 | 8.1 | 0.205 |
| | WBGT(°C) | 34.2 | 28.6 | 29.7 | 31.4 | 39.3 | 48.5 | 5.3 | 0.155 |
| Pedestrian-level | PET(°C) | 37.7 | 29.9 | 34.1 | 35.3 | 39.7 | 59.6 | 5.2 | 0.138 |
| | WBGT(°C) | 32.7 | 28.9 | 29.7 | 31.0 | 35.0 | 44.6 | 3.8 | 0.116 |
| Underground | PET(°C) | 37.5 | 34.8 | 37.2 | 37.4 | 37.8 | 38.7 | 0.5 | 0.013 |
| | SET(°C) | 35.6 | 32.7 | 35.3 | 35.6 | 36.0 | 36.7 | 0.5 | 0.014 |
| | WBGT(°C) | 32.9 | 30.8 | 32.7 | 32.9 | 33.1 | 34.1 | 0.4 | 0.012 |

Table B4: Descriptive statistics of thermal comfort indices at site L2

| Site L2 | Variable | Mean | Min | 25% | 50% | 75% | Max | Std | CV |
|---|---|---|---|---|---|---|---|---|---|
| Roof | PET(°C) | 40.6 | 29.1 | 33.6 | 37.0 | 47.8 | 64.3 | 8.5 | 0.209 |
| | WBGT(°C) | 35.1 | 28.4 | 30.1 | 33.3 | 40.6 | 48.6 | 5.6 | 0.160 |
| Pedestrian-level | PET(°C) | 36.6 | 28.9 | 33.4 | 35.9 | 38.6 | 62.9 | 4.6 | 0.126 |
| | WBGT(°C) | 32.6 | 28.5 | 29.4 | 31.8 | 34.9 | 47.3 | 3.5 | 0.107 |
| Underground | PET(°C) | 33.9 | 29.0 | 33.6 | 33.9 | 34.2 | 35.6 | 0.5 | 0.015 |
| | SET(°C) | 32.3 | 26.7 | 31.8 | 32.3 | 32.7 | 34.1 | 0.6 | 0.019 |
| | WBGT(°C) | 29.3 | 27.6 | 29.1 | 29.3 | 29.5 | 30.8 | 0.4 | 0.014 |

Table B5: Kolmogorov–Smirnov test of environmental variables measured at site L1

| Variable | Dataset #1 | Dataset #2 | KS Statistic | p-value |
|---|---|---|---|---|
| AT (°C) | Roof | L1 Pedestrian-level | 0.125 | 6.21E-09 |
| | Roof | L1 Underground | 0.512 | 7.78E-149 |
| | L1 Pedestrian-level | L1 Underground | 0.552 | 1.46E-174 |
| WBT (°C) | Roof | L1 Pedestrian-level | 0.187 | 1.82E-19 |
| | Roof | L1 Underground | 0.480 | 2.05E-130 |
| | L1 Pedestrian-level | L1 Underground | 0.496 | 3.26E-139 |
| GT (°C) | Roof | L1 Pedestrian-level | 0.233 | 1.56E-29 |
| | Roof | L1 Underground | 0.437 | 2.00E-105 |
| | L1 Pedestrian-level | L1 Underground | 0.485 | 1.36E-132 |
| MRT (°C) | Roof | L1 Pedestrian-level | 0.201 | 4.43E-22 |
| | Roof | L1 Underground | 0.464 | 4.27E-119 |
| | L1 Pedestrian-level | L1 Underground | 0.480 | 7.56E-130 |
| V(m/s) | Roof | L1 Pedestrian-level | 0.198 | 1.15E-21 |
| | Roof | L1 Underground | 0.598 | 5.83E-207 |
| | L1 Pedestrian-level | L1 Underground | 0.400 | 2.28E-89 |
| RH (%) | Roof | L1 Pedestrian-level | 0.162 | 1.02E-14 |
| | Roof | L1 Underground | 0.579 | 1.76E-193 |
| | L1 Pedestrian-level | L1 Underground | 0.644 | 1.49E-243 |

Table B6: Kolmogorov–Smirnov test of environmental variables measured at site L2

| Variable | Dataset #1 | Dataset #2 | KS Statistic | p-value |
|---|---|---|---|---|
| AT (°C) | Roof | L2 Pedestrian-level | 0.223 | 5.50E-39 |
| | Roof | L2 Underground | 0.699 | 0 |
| | L2 Pedestrian-level | L2 Underground | 0.640 | 0 |
| WBT (°C) | Roof | L2 Pedestrian-level | 0.265 | 5.75E-55 |
| | Roof | L2 Underground | 0.445 | 8.95E-159 |
| | L2 Pedestrian-level | L2 Underground | 0.385 | 9.58E-117 |
| GT (°C) | Roof | L2 Pedestrian-level | 0.284 | 5.21E-63 |
| | Roof | L2 Underground | 0.703 | 0 |
| | L2 Pedestrian-level | L2 Underground | 0.658 | 0 |
| MRT (°C) | Roof | L2 Pedestrian-level | 0.250 | 1.18E-48 |
| | Roof | L2 Underground | 0.699 | 0 |
| | L2 Pedestrian-level | L2 Underground | 0.681 | 0 |
| V(m/s) | Roof | L2 Pedestrian-level | 0.188 | 1.23E-27 |
| | Roof | L2 Underground | 0.442 | 2.65E-156 |
| | L2 Pedestrian-level | L2 Underground | 0.437 | 9.11E-151 |
| RH (%) | Roof | L2 Pedestrian-level | 0.228 | 1.59E-40 |
| | Roof | L2 Underground | 0.955 | 0 |
| | L2 Pedestrian-level | L2 Underground | 0.835 | 0 |

Table B7: Kolmogorov–Smirnov test of environmental variables between site L1 and site l2

| Variable | Location | KS Statistic | p-value |
|---|---|---|---|
| AT (°C) | Pedestrian-level | 0.15304 | 7.28e-22 |
| | Underground | 0.998477 | 0.00e+00 |
| WBT (°C) | Pedestrian-level | 0.301126 | 2.67e-84 |
| | Underground | 0.863308 | 0.00e+00 |
| GT (°C) | Pedestrian-level | 0.109716 | 1.96e-11 |
| | Underground | 0.997335 | 0.00e+00 |
| MRT (°C) | Pedestrian-level | 0.137266 | 1.15e-17 |
| | Underground | 0.99677 | 0.00e+00 |
| V(m/s) | Pedestrian-level | 0.205334 | 4.07e-39 |
| | Underground | 0.000566 | 1.00e+00 |
| RH (%) | Pedestrian-level | 0.230886 | 1.98e-49 |
| | Underground | 0.99095 | 0.00e+00 |

Table B8: One-way ANOVA on effect of MET on PET at different locations

| Location | Source | Sum of Squares | df | F-statistic | p-value |
|---|---|---|---|---|---|
| Roof | MET | 291549.94 | 8 | 2237829 | 0 |
| | Residual | 732.69 | 44,991 | | |
| Pedestrian-level | MET | 262413.10 | 8 | 5510150 | 0 |
| | Residual | 267.83 | 44,991 | | |
| Underground | MET | 215348.19 | 8 | 50124000 | 0 |
| | Residual | 24.16 | 44,991 | | |